\documentclass[prd,aps,preprint,tightenlines,nofootinbib,superscriptaddress]{revtex4}

\usepackage{amsmath}
\usepackage{epsfig}
\usepackage{axodraw}

\begin{document}

\hfill{NPAC-09-16}

\title{A Comprehensive Analysis of Electric Dipole Moment Constraints on CP-violating Phases in the MSSM}

\author{Yingchuan Li}
\email{yli@physics.wisc.edu} \affiliation{Department of Physics,
University of Wisconsin, Madison, Wisconsin 53706 USA}
\author{Stefano Profumo }
\email{profumo@scipp.ucsc.edu} \affiliation{Department of Physics
and Santa Cruz Institute for Particle Physics, University of
California, 1156 High St., Santa Cruz, CA 95064, USA}
\author{Michael Ramsey-Musolf}
\email{mjrm@physics.wisc.edu} \affiliation{Department of Physics,
University of Wisconsin, Madison, Wisconsin 53706 USA}
\affiliation{Kellogg Radiation Laboratory, California Institute of
Technology, Pasadena, CA 91125 USA}

\date{\today}

\begin{abstract}

\noindent We analyze the constraints placed on individual, flavor diagonal
CP-violating phases in the minimal supersymmetric extension of the
Standard Model (MSSM)  by current experimental bounds
on the electric dipole moments (EDMs) of the neutron,  Thallium,
and Mercury atoms.
We identify the four CP-violating phases that are individually highly constrained by current EDM bounds, 
and we explore how these phases and correlations among them are constrained by current EDM limits. We also analyze the prospective implications of  the next 
generation of EDM experiments. We point out that all other CP-violating phases in the MSSM are not nearly as tightly constrained by limits on the size of EDMs.
We emphasize that a rich set of phenomenological consequences is potentially
associated with these generically large EDM-allowed phases, ranging
from B physics, electroweak baryogenesis, and signals of
CP-violation at the CERN Large Hadron Collider and at future linear colliders. Our numerical
study takes into account the complete set of contributions
from one- and two-loop EDMs of the electron and quarks, one- and
two-loop Chromo-EDMs of quarks, the Weinberg 3-gluon operator, and
dominant 4-fermion CP-odd operator contributions, including
contributions which are both included and not included yet in the
CPsuperH2.0 package. We also introduce an open-source numerical
package, {\tt 2LEDM}, which provides the complete set of two-loop
electroweak diagrams contributing to the electric dipole moments of
leptons and quarks.
 \end{abstract}

\maketitle

\section{introduction}

The minimal supersymmetric extension of the standard model (MSSM)
features a large number of additional CP-violating sources compared to the SM, yielding an
extremely rich array of possible phenomenological consequences
\cite{Ibrahim:2007fb,RamseyMusolf:2006vr}. To mention a few, the new
sources of CP violation can participate in the generation of the
baryon asymmetry of the universe in the context of supersymmetric
electroweak baryogenesis \cite{EWB,Li:2008ez}. They can generate
observable CP asymmetries at the CERN Large Hadron Collider (LHC)
\cite{Bartl:2004jr,Langacker:2007ur,Ellis:2008hq} as well as at
future linear colliders
\cite{Choi:1999cc,Barger:2001nu,Kizukuri:1990iy,Bartl:2003tr,Bartl:2004jj,CPVILCtau}.
They could manifest themselves in B physics
(see Ref.~\cite{Bphysics,flavorblind,Kane:2003zi} and the recent results from the Tevatron D0 Collaboration \cite{tevatronCP} and CDF Collaboration\cite{CDFCP}). In general, CP-violating phases
induce rather large contributions, compared to current experimental
sensitivity, to the electric dipole moment (EDM) of the neutron and
of atoms \cite{Pospelov:2005pr}. A large CP-violating phase could
also have an effect on many CP-conserving observables such as the
mass spectrum, the production rate and decay branching ratios of
SUSY particles, especially the lightest neutralino as a dark matter
particle (including direct and indirect detection rates, and the
relic density) \cite{CPCONSERVING,phi1DM,phitbDM} and Supersymmetric Higgs scalars\cite{Pilaftsis:1999qt,Carena:2002bb}.

All these potentially interesting observable signatures are however
highly constrained by the fact that no  permanent EDM has
ever been experimentally observed. The current most stringent bounds
are on the EDMs of the neutron \cite{Baker:2006ts}, and on that of
Thallium \cite{Regan:2002ta} and Mercury \cite{Griffith:2009zz}
atoms\footnote{For recent reviews on EDM searches and their
implications for the MSSM, see, {\em e.g.}
Refs.~\cite{Pospelov:2005pr,RamseyMusolf:2006vr,Ellis:2008zy}.}.
Specifically, the current experimental constraints on the
aforementioned quantities read:
\begin{eqnarray}
|d_{{\rm n}}| &<& 2.9 \times 10^{-26} e ~{\rm cm} ~(90\% {\rm C.L.}), \nonumber \\
|d_{{\rm Tl}}| &<& 9.0 \times 10^{-25} e ~{\rm cm} ~(90\% {\rm C.L.}), \nonumber \\
|d_{{\rm Hg}}| &<& 3.1 \times 10^{-29} e ~{\rm cm} ~(95\% {\rm
C.L.}).
\end{eqnarray}

Signatures of CP-violation compatible with the bounds described
above have so far been discussed in the context of scenarios where
one avoids EDM bounds and keeps relatively large phases with light
super-particle masses via one or more cancelations among the various
terms contributing to the EDMs
\cite{Falk:1996ni,Ibrahim:1997nc,Ibrahim:1997gj,Ibrahim:1998je,Brhlik:1998zn,Barger:2001nu,Ibrahim:2000qj,Ellis:2008zy}.
This ``cancellation'' scenario occurs typically in some very
fine-tuned region\footnote{It is estimated in Ref.
\cite{Abel:2001vy} that a minimal degree of fine-tuning at the level of $10^{-2}$
is needed.} of the MSSM parameter space. The goal of the present
analysis is instead to examine in detail, beyond the
possibility of cancelations between different CP-violating
sources\footnote{There are still cancelations among different
contributions to EDMs from the same source in some region of
parameter space. We emphasize that our main focus is on the general
trend of constraints without cancelation, and always point it out
when we encounter such cancelation region.}, how each individual
phase is constrained by current EDM bounds and to study under which
conditions large phases are phenomenologically allowed while also
keeping the relevant mass scales relatively light and thus phenomenologically interesting. Instead of
imposing any universality condition as in the supergravity (SUGRA)
model, we entertain here the possibility of general non-universal soft terms (as is the case in many string-inspired
models, see e.g. \cite{nonuniversal}), and thus keep all the soft terms in MSSM as
independent variables. For concreteness, we focus on flavor diagonal phases. We defer a study of EDM bounds on flavor non-diagonal phases  to future work.
One of our main results is that among all MSSM CP-violating phases, only four
are in fact strongly constrained by current EDM bounds, and we study
the correlations among them. We comment on the possible phenomenological implications of those
phases that are not strongly constrained. 

Part of our numerical study is based on the CPsuperH2.0 package
\cite{Lee:2003nta,Lee:2007gn}, which includes the complete
contributions from one-loop supersymmetric contributions to the EDMs
of the electron and of quarks, one-loop and two-loop Chromo-EDMs of
quarks \cite{Ibrahim:1997nc,Ibrahim:1997gj,Ibrahim:1998je,2loop},
6-dimensional 3-gluon Weinberg operator \cite{3gluon}, and dominant
contributions to 4-fermion CP-odd operators \cite{Demir:2003js}, but
only a subset of the contributions of two-loop EDMs of electron and
quarks \cite{2loop}. There exist additional  Higgs exchange-mediated
chargino-neutralino 2-loop contributions to electron and quark EDMs that can become dominant in the limit of heavy sfermions
\cite{Li:2008kz} and that are not included in CPsuperH2.0 package. We
here take into account these contributions as well, therefore featuring the
dominant 1-loop and 2-loop contributions to all the 6-dimensional
CP-odd operators that generate EDMs of the neutron and of Thallium
and Mercury atoms\footnote{The four fermion operators in the MSSM are technically dimension eight.}. The numerical code where all these new contributions are collected is called {\tt 2LEDM}, and is currently available from the authors upon request. The  {\tt 2LEDM} code currently includes an interface to the FeynHiggs\footnote{\tt http://www.feynhiggs.de/} package (version 2.6.5) \cite{Hahn:2009zz}. We plan in the near future to set up a webpage for easier download of  {\tt 2LEDM} and of related tutorials and to include an interface to the CPsuperH2.0 package.

There remains an order one theoretical uncertainty
with neutron and Mercury EDMs. For the neutron EDM, the uncertainty arises
from hadronic physics, while for the Mercury EDM the source of uncertainty is
associated with (i) atomic physics in extracting the nuclear Schiff moment from
$d_{{\rm Hg}}$, (ii) the nuclear physics going into extracting T- and P-odd pion-nucleon
couplings $\bar{g}_{\pi NN}$ from the Schiff moment, and (iii) the hadronic physics
in computing the $\bar{g}_{\pi NN}$ in terms of quark Chromo-EDM
operator, Weinberg three-gluon operator, and CP-violating four fermion operators\footnote{In the MSSM, the Chromo-EDM operator typically gives the dominant contribution to the $\bar{g}_{\pi NN}$ .}. In utilizing the CPsuperH2.0 package to estimate the relevant EDMs from the EDMs of quarks and leptons we are relying  on QCD sum rule computations
\cite{Demir:2003js,QCDsumrule,Olive:2005ru} of strong interaction matrix elements. For a discussion of the systematic uncertainties in our results for these quantities when different hadronic model approximations are employed we refere the reader to e.g. Ref.~\cite{Ellis:2008zy}. We note also that the CPsuperH2.0 code relies on the computations of the nuclear Schiff moment reported in Ref.~\cite{Dmitriev:2005id} and does not take into account the recent computations of Refs.~\cite{deJesus:2005nb,Ban:2010ea}. The latter two computations give an enhanced sensitivity of the nuclear Schiff moment of $^{199}$Hg to the isoscalar, T- and P-odd pion-nucleon coupling, ${\bar g}_\pi^{(0)}$ as compared to Ref.~\cite{Dmitriev:2005id}. %We do not expect this difference to have a substantial bearing on the present analysis, as the Schiff moment is roughly ten times more sensitive to the isovector coupling, ${\bar g}_\pi^{(1)}$ than to ${\bar g}_\pi^{(0)}$. 
Moreover, Ref.~\cite{Ban:2010ea} finds that the sensitivity to ${\bar g}_\pi^{(1)}$ may be reduced by a factor of three to ten, depending on the type of interaction used. 

In general,  ${\bar g}_\pi^{(j)}$ are dominated by quark Chromo-EDMs rather than the Weinberg three gluon operator (the contribution is suppressed by $m_q$), while in the MSSM with large $\tan\beta$, contributions from the four fermion operators may be important\cite{Pospelov:2005pr,Lebedev:2002ne}. The QCD sum rule analysis implies that ${\bar g}_\pi^{(0)}$ is five times less sensitive to the sum of up and down quark Chromo-EDMs than ${\bar g}_\pi^{(1)}$ is to their difference. Thus, we would only expect the stronger sensitivity of the Mercury Schiff moment to ${\bar g}_\pi^{(0)}$ to be important in small corners of the MSSM parameter space where the difference of quark Chromo-EDMs is highly suppressed compared to their average value. The possible suppression in sensitivity to ${\bar g}_\pi^{(1)}$ is a potentially more serious issue.
Thus, one may need to relax the constraints we obtain on CP-violating phases that are driven by the $^{199}$Hg results in light of on-going theoretical nuclear structure developments.

With these caveats in mind, we summarize our main findings here:
\begin{itemize}
\item[(a)] A primary impact of the new $^{199}$Hg result is to impose significantly more stringent constraints on the relative phase $\phi_3$ between the gluino soft supersymmetric-breaking mass and the $\mu$ parameter (see below), while generating a strong correlation between this phase and the phase of the soft-breaking triscalar couplings involving first generation sfermions.
\item[(b)] The neutron and Thallium EDM limits have a stronger impact on the relative phase $\phi_2$ between the wino soft mass parameter and $\mu$ than does the $^{199}$Hg bound, but at present there does not exist any strong correlation between $\phi_2$ and other phases. 
\item[(c)] A future neutron EDM limit that is roughly 100 times stronger than present would both tighten the present correlations between $\phi_3$ and the triscalar phases while inducing strong correlations between $\phi_2$ and other phases.
\item[(d)] In the limit of heavy first and second generation sfermions, the \lq\lq bino" phase $\phi_1$ is essentially unconstrained by present EDM bounds. A future neutron or electron EDM measurement with $\sim 100$ times better sensitivity would probe the impact of this phase at a level of interest for cosmology.
\end{itemize}

In the remainder of the paper, we organize the discussion of our analysis leading to the findings above as follows: In section \ref{sec:general}, we
give a general discussion about the CP-violating phase structure of
the MSSM, we address how each phase impacts the various EDMs, and 
we outline the eneral setup of our analysis. In section \ref{sec:detail}, we
investigate in detail how each phase is constrained by current EDM
bounds, we study the correlations between the various EDM bounds on the most strongly
constrained phases, namely $\phi_2$, $\phi_3$, and $\phi_{u,d}$, and we
discuss the phenomenology implications of the other loosely
constrained phases. Finally, we devote section \ref{sec:conclusions} to
our summary and conclusions.

\section{CP-violating phases in MSSM and the setup for analysis}
\label{sec:general}

\begin{table}[!b]
\caption{Summary of how the CP-violating sources in MSSM generate
various CP-odd operators at one-loop and two-loop level.}
\begin{center}
\label{tab:phase2operator}
\begin{tabular}{|c|c|c|}
\hline CP-violating phases & one-loop contribution &
two-loop contribution \\
\hline $\phi_{e,u,d}$ &
$d^{1-loop}_{u,d,e}$,$\tilde{d}^{1-loop}_{u,d}$, $C_{ff'}$
& no  \\
\hline $\phi_{\mu,c,s}$ & no & no \\
\hline $\phi_{\tau,t,b}$ & no &
$d^{2-loop}_{u,d,e}(\tilde{t},\tilde{b},\tilde{\tau})$,
$\tilde{d}^{2-loop}_{u,d}(\tilde{t},\tilde{b},\tilde{\tau})$, $d^{{\rm 3G}}$ \\
\hline $\phi_{1,2}$ &
$d^{1-loop}_{u,d,e}$,$\tilde{d}^{1-loop}_{u,d}$,
$C_{ff'}$ & $d^{2-loop}_{u,d,e}(\chi^{\pm,0})$ \\
\hline $\phi_3$ & $d^{1-loop}_{u,d}$,$\tilde{d}^{1-loop}_{u,d}$,
$C_{ff'}$ & $d^{3{\rm G}}$ \\
\hline
\end{tabular}
\end{center}
\end{table}

\begin{table}[!b]
\caption{Summary of relevant CP-odd operators of neutron, Thallium,
and Mercury atom EDMs.}
\begin{center}
\label{tab:operator2edm}
\begin{tabular}{|c|c|c|}
\hline $d_n$ &
$d_{\rm Tl}$ & $d_{\rm Hg}$ \\
\hline $d_{u,d}$, $\tilde{d}_{u,d}$,  $d^{3G}$, $C_{ff'}$
& $d_e$, $C_{ff'}$ & $d_e$, $\tilde{d}_{u,d}$, $C_{ff'}$\\
\hline
\end{tabular}
\end{center}
\end{table}
The Minimal Supersymmetric Extension to the Standard Model of particle physics introduces a plethora of new and unknown parameters. Many of
these parameters are connected to new sources of CP or flavor
violation, or both. Although EDMs could, in principle, be induced by
all CP-violating parameters including both flavor-conserving and
flavor-violating ones, they are most sensitive to flavor-conserving
CP-violating phases, including those associated with the bilinear
coupling $b$ and Higgsino mass term $\mu$ in the Higgs-Higgsino
sector,  the soft-supersymmetry breaking Majorana masses $M_1$,
$M_2$, and $M_3$ in the gaugino sector, and  the trilinear
couplings $A_f$ in the sfermion sector.

This notwithstanding, not all the new CP violating phases appearing in the MSSM
are physical. In fact, there exist two transformations that can be
employed to rotate away two phases \cite{Chung:2003fi}. We choose a
convention where $\mu$ and $b$ are real, and the remaining phases
mentioned above are all physical. In particular, the physical phases
include the phases $\phi_{1,2,3}$ of the gaugino masses $M_{1,2,3}$,
and the phases $\phi_{u,d,e}$, $\phi_{c,s,\mu}$, and
$\phi_{t,b,\tau}$ of the sfermion trilinear couplings $A_{u,d,e}$,
$A_{c,s,\mu}$, and $A_{t,b,\tau}$, respectively. As shown in Table
\ref{tab:phase2operator}, these phases play different roles in
generating various CP-odd operators, including the electron EDM
$d_e$, quark EDM $d_q$ and Chromo-EDM $\tilde{d}_q$, the 
Weinberg 3-gluon operator $d^{3G}$, and the 4-fermion CP-odd operator
$C_{ff'}$\footnote{For the specific form of each of these operators,
 see, {\em e.g.} Refs.~\cite{Ellis:2008zy}.}. These CP-odd operators are responsible for
the EDMs of the neutron, as well as of that of the Thallium and
Mercury atoms, as summarized in Table \ref{tab:operator2edm}. In
particular, in the MSSM the Thallium EDM is dominated by the electron EDM
operator $d_e$, and possibly by the four-fermion operator $C_{ff'}$
if ${\rm tan}\beta > 30$ \cite{Demir:2003js}; the neutron EDM, which
we compute here using QCD sum rule results \cite{QCDsumrule},
mainly stems from the EDM and chromo-EDM operators of the $u$ and
$d$ quarks, $d_{u,d}$ and $\tilde{d}_{u,d}$, and from the 3-gluon
term $d^{3G}$; lastly, the Mercury EDM is generated primarily by the
chromo-EDM operators $\tilde{d}_{u,d}$ \cite{Pospelov:2005pr}. A
combination of Table \ref{tab:phase2operator} and Table
\ref{tab:operator2edm} provides information on how each CP-violating
phase is constrained by which experimental EDM bound.

Among all contributions, some of the dominant ones stem from the
one-loop induced EDM and Chromo-EDM operators $d_e$, $d_{u,d}$, and
$\tilde{d}_{u,d}$. These contributions always involve the first-two
generations of sleptons and squarks, and therefore are
asymptotically suppressed in the limit where these scalar fermions are very heavy
\cite{Cohen:1996vb}. Obviously, the effect of the CP-violating
phases $\phi_u,\phi_d,\phi_s,\phi_c,\phi_e,\phi_\mu$ from the
first-two generations sfermions are completely suppressed
in this situation, and would not show up in any other observable
signature. In contrast, the effects of other phases, including
$\phi_{1,2,3}$ in the gaugino sector and $\phi_t,\phi_b,\phi_\tau$
in the third-generation sfermion sector (thanks to larger Yukawa couplings), are not as strongly suppressed in the decoupling limit of heavy first and second generation sfermions, and they might induce interesting effects that could manifest
themselves at colliders or in other experiments sensitive to CP
violation. 

With these considerations in mind, we study cases where the sfermion masses for the first
two generations are either light or heavy. For each case, we explore in
detail the mass- and ${\rm tan} \beta$-dependence of the EDM
bounds on each individual phase. We choose a set of mass parameters
corresponding to a light spectrum as the reference point (we call this ``Case I''); we then
study the effect on the constraints as the relevant
mass scales increase (the limiting case is indicated as ``Case II''). The reference values we choose for the relevant supersymmetric
parameters $M_{1,2,3}$, $\mu$, $A_f$ of all flavors $f$, the charged
Higgs mass $M_{H^{\pm}}$, and the third generation sfermion masses
$m_{L_3,R_3}$ in both cases are as follows:
\begin{eqnarray} && |M_1| = 150~{\rm GeV}, |M_2| = 250~{\rm
GeV}, |M_3| = 550~{\rm GeV},  \nonumber \\
&& |\mu| = 225~{\rm GeV}, |A_f| = 175~{\rm GeV}, M_{H^{\pm}}=500{\rm GeV}, \nonumber \\
&& m_{L_3} = m_{R_3} = 200~{\rm GeV},
\end{eqnarray}
We set the first-two generation sfermion masses, in the
two cases, to:
\begin{eqnarray}
&& {\rm CASE~ I: } ~ m_{L_{1,2}} = m_{R_{1,2}} = 200~{\rm GeV}, \label{eq:case1} \\
&& {\rm CASE~ II: } ~ m_{L_{1,2}} = m_{R_{1,2}} = 10~{\rm TeV}.
\label{eq:case2}
\end{eqnarray}
In the study of each phase, we look at how EDM constraints are affected by changes in the relevant mass scale, keeping all other masses set to their reference values.

\begin{figure}
\begin{center}
\mbox{\epsfig{file=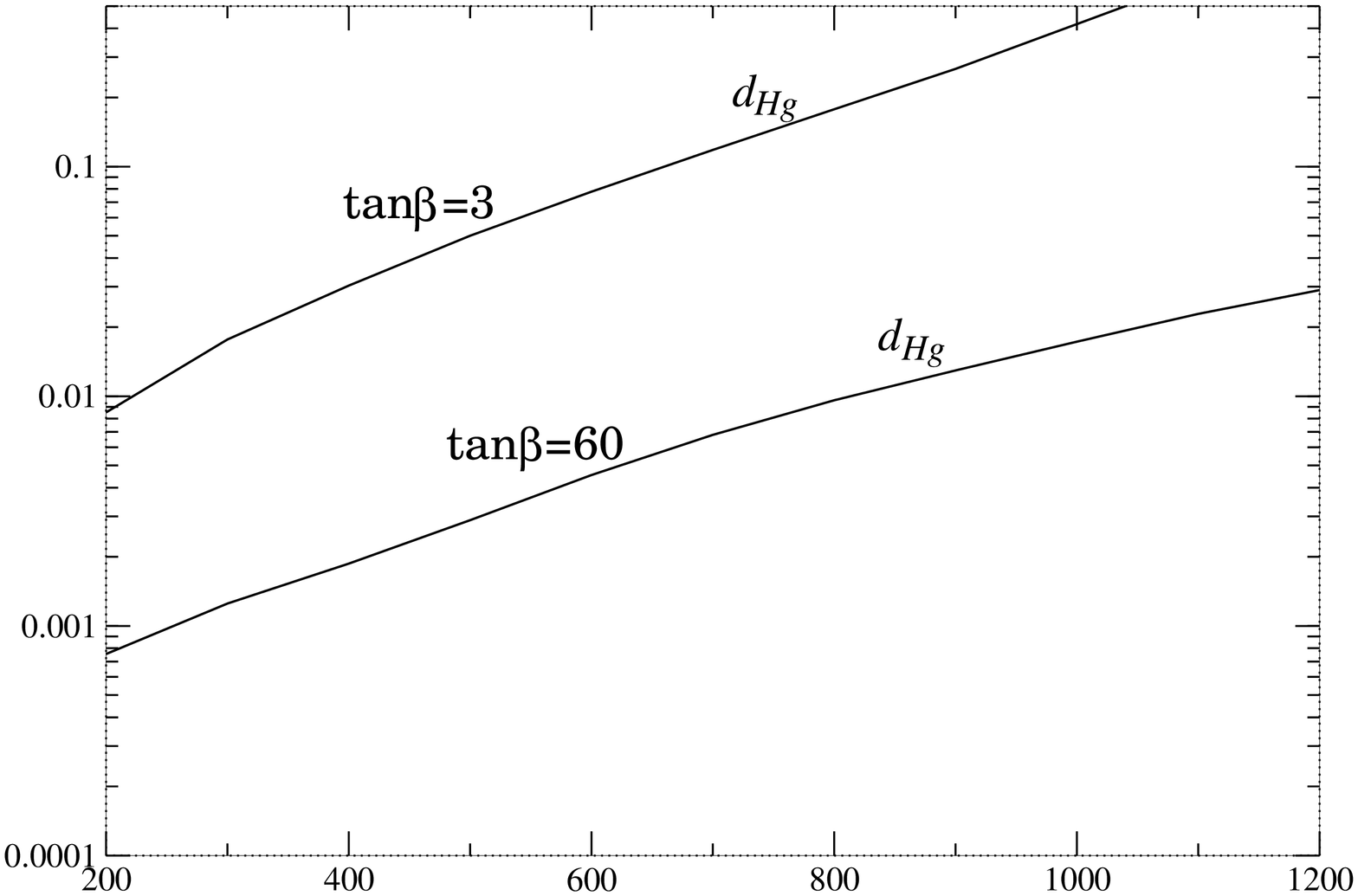,width=6.75cm,angle=-90}
\epsfig{file=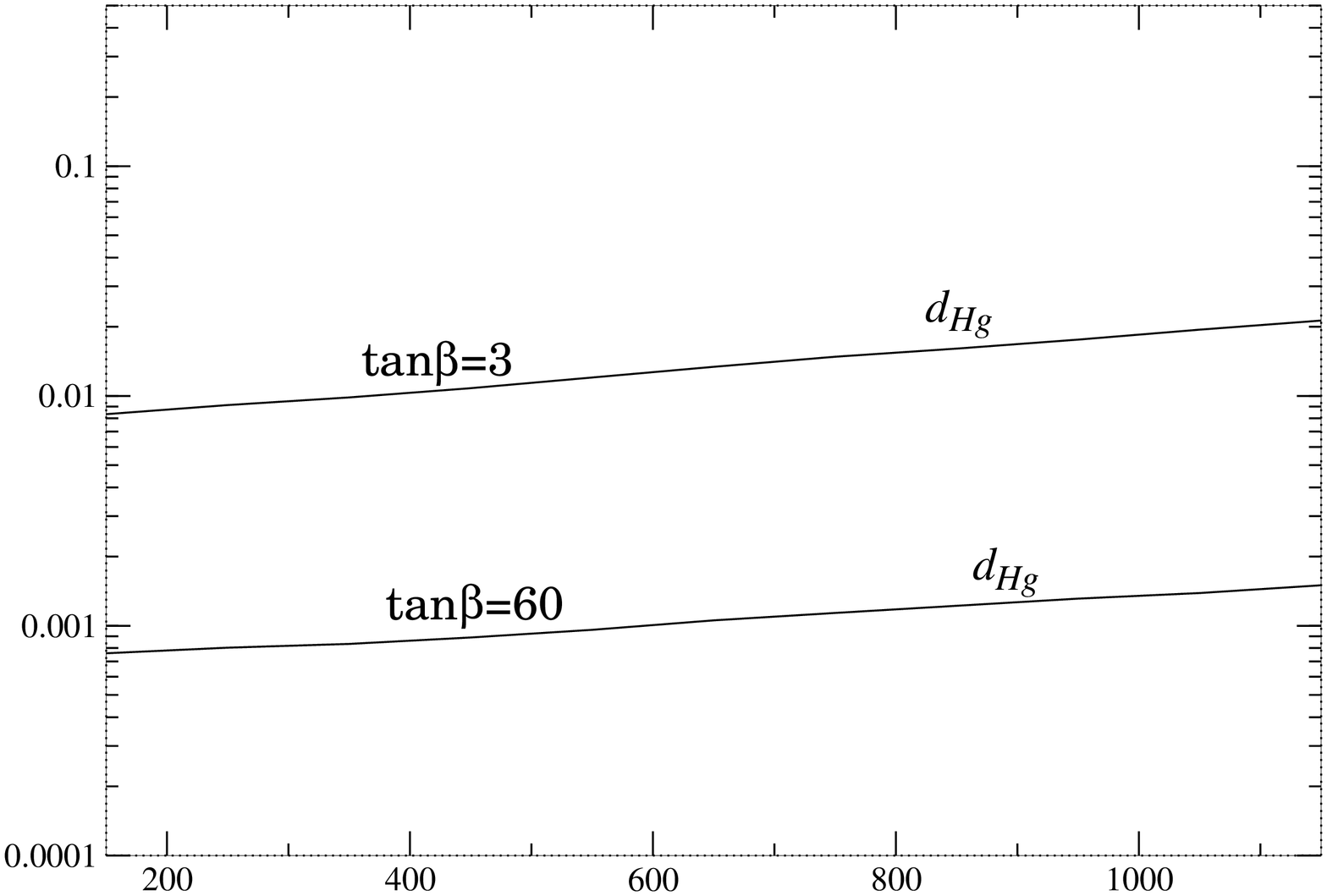,width=6.75cm,angle=-90}}
\put(-410,-185){$(m_{L,R})_1$ [GeV]} \put(-140,-185){$M_1$ [GeV]}
\put(-500,-40){$\phi_1(\pi)$} \put(-248,-40){$\phi_1(\pi)$}\
\mbox{\epsfig{file=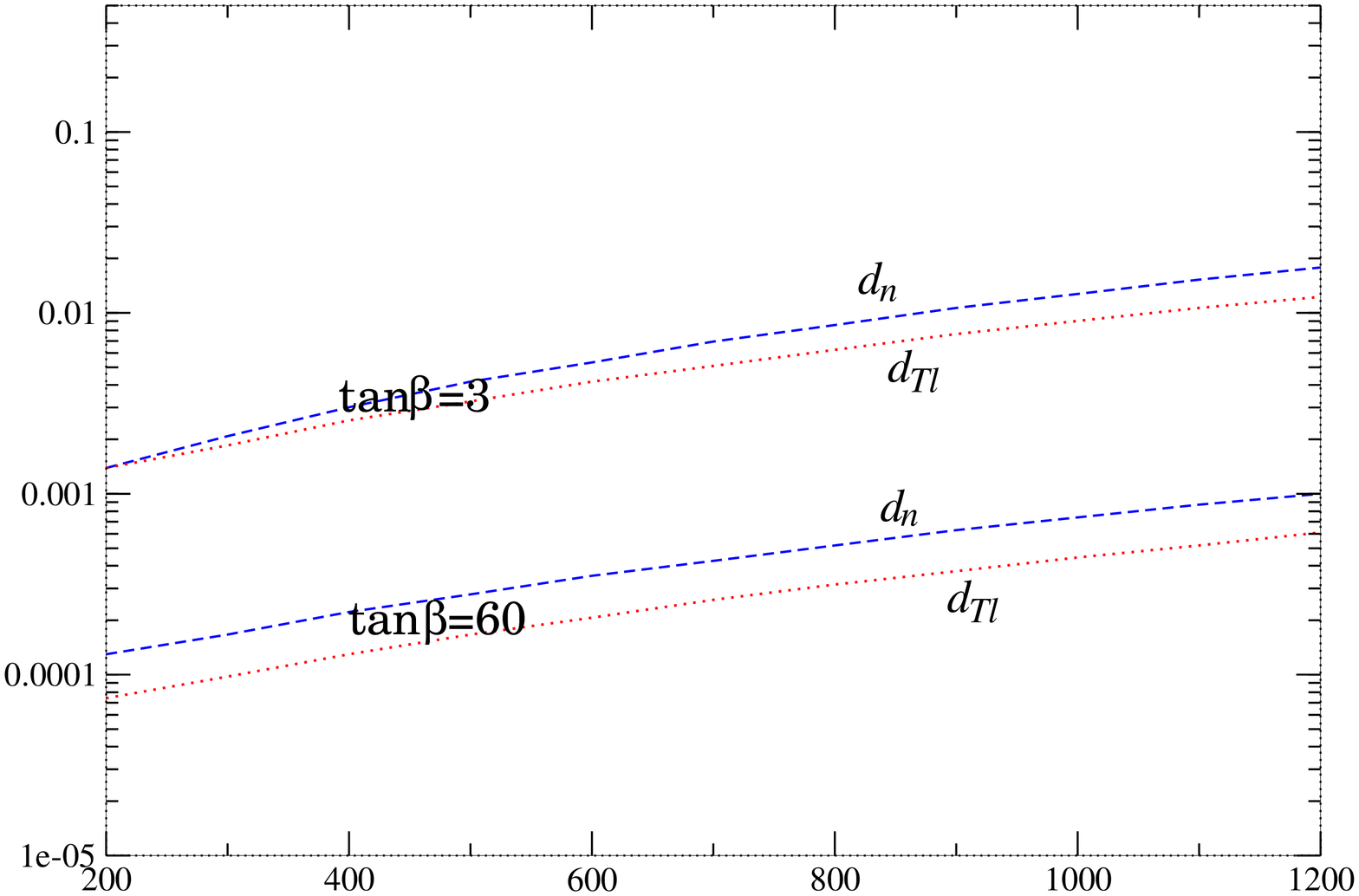,width=6.75cm,angle=-90}
\epsfig{file=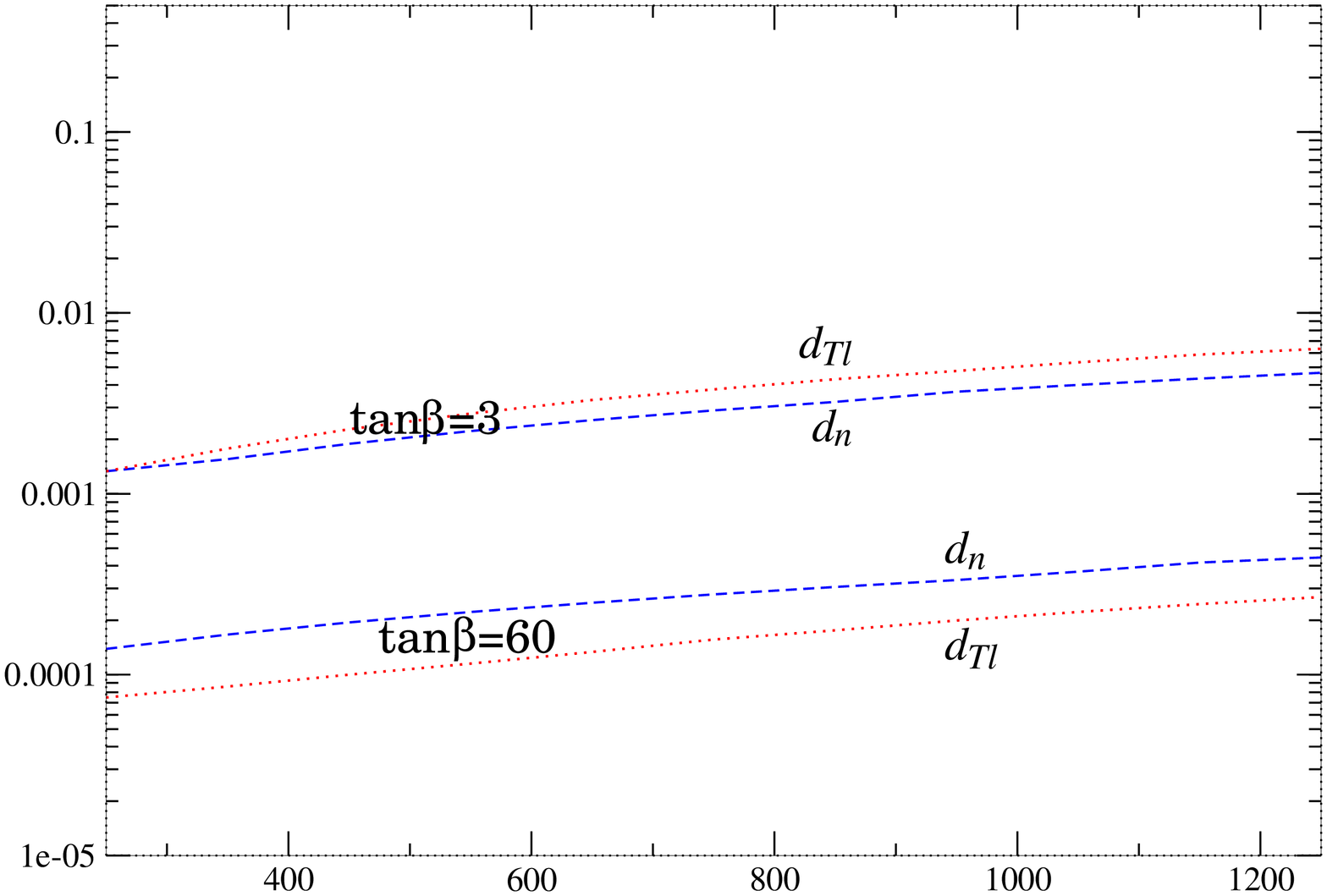,width=6.75cm,angle=-90}}
\put(-410,-185){$(m_{L,R})_1$ [GeV]} \put(-140,-185){$M_2$ [GeV]}
\put(-500,-40){$\phi_2(\pi)$} \put(-248,-40){$\phi_2(\pi)$}\\ 
\mbox{\epsfig{file=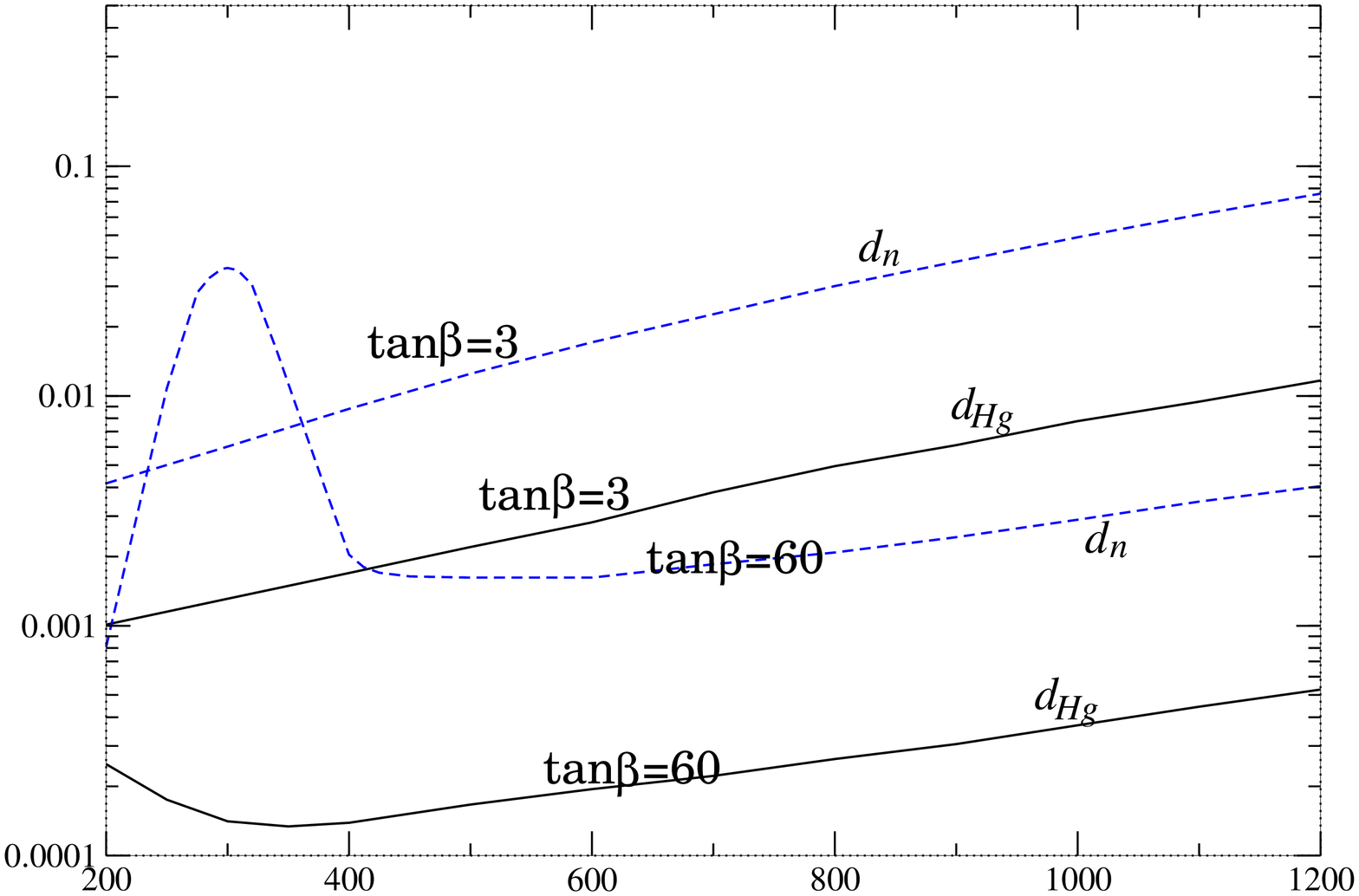,width=6.75cm,angle=-90}
\epsfig{file=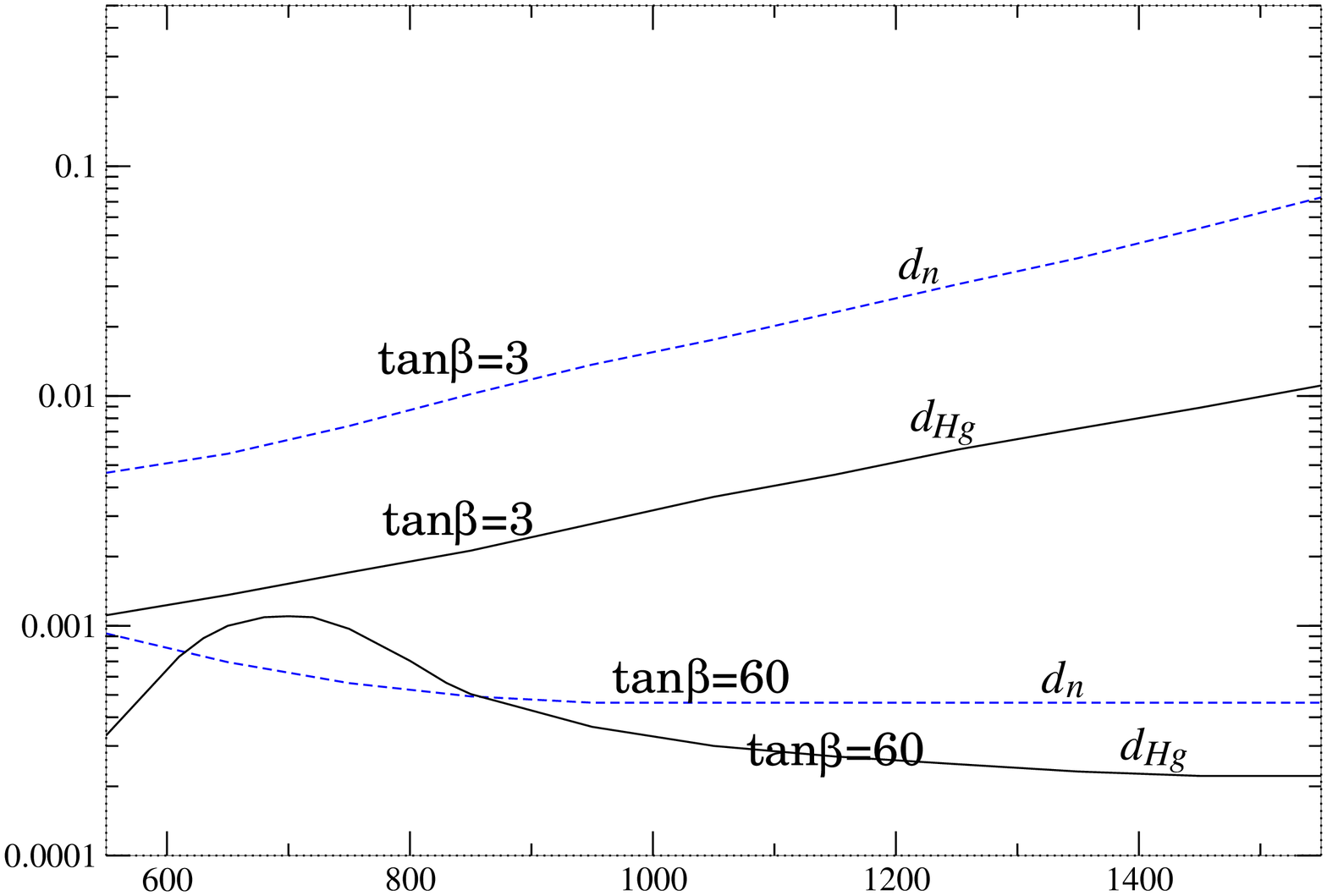,width=6.75cm,angle=-90}}
\put(-410,-185){$(m_{L,R})_1$ [GeV]} \put(-140,-185){$M_3$ [GeV]}
\put(-500,-40){$\phi_3(\pi)$} \put(-248,-40){$\phi_3(\pi)$} 
\end{center}

\caption{Constraints on the CP violating phases $\phi_1$, $\phi_2$ and $\phi_3$ (from top to bottom) versus $(m_{L,R})_1$ (left panels) and versus $M_{1,\ 2,\ 3}$ (right panels) from experimental limits on the Mercury (black solid lines), Tallium (red dotted) and neutron (blue dashed) EDMs, in the case of light first-two generations
of sfermions ({\em case I}, see Eq. \ref{eq:case1}).}
\label{fig:caseI}
\end{figure}

\begin{figure}
\centerline{\epsfig{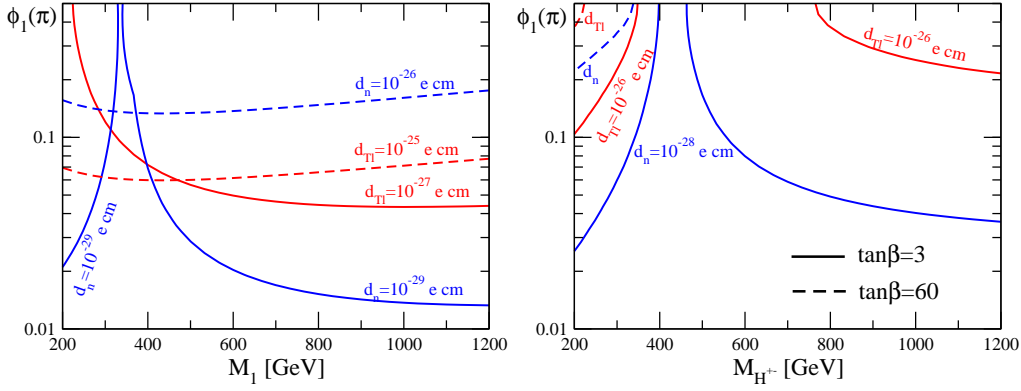}}
%\epsfig{file=phi1_M1_heavy.ps,width=6.75cm,angle=-90}
%\epsfig{file=phi1_MA_heavy.ps,width=6.75cm,angle=-90}
%\put(-410,-185){$M_1$ [GeV]} \put(-140,-185){$M_{H^{\pm}}$ [GeV]}
%\put(-500,-40){$\phi_1(\pi)$} \put(-248,-40){$\phi_1(\pi)$} }
\caption{Curves of constant values for the Thallium (red) and neutron (blue) EDM
as a function of $M_1$ and $M_{H^{\pm}}$ in the case of heavy
first-two generations of sfermions ({\em case II}) as in Eq.
\ref{eq:case2}.  Because current EDM limits do not constrain $\phi_1$ in this case, curves correspond to representative
future EDM sensitivities. Solid and dashed curves correspond, respectively, to $\tan\beta=3$ and 60. 
%$d_{Tl}=5.9\times 10^{-28}$ e cm corresponds to an
%electron EDM value of $d_e=10^{-30}$ e cm.
} \label{fig:phi1b}
\end{figure}

\begin{figure}
\centerline{
\epsfig{file=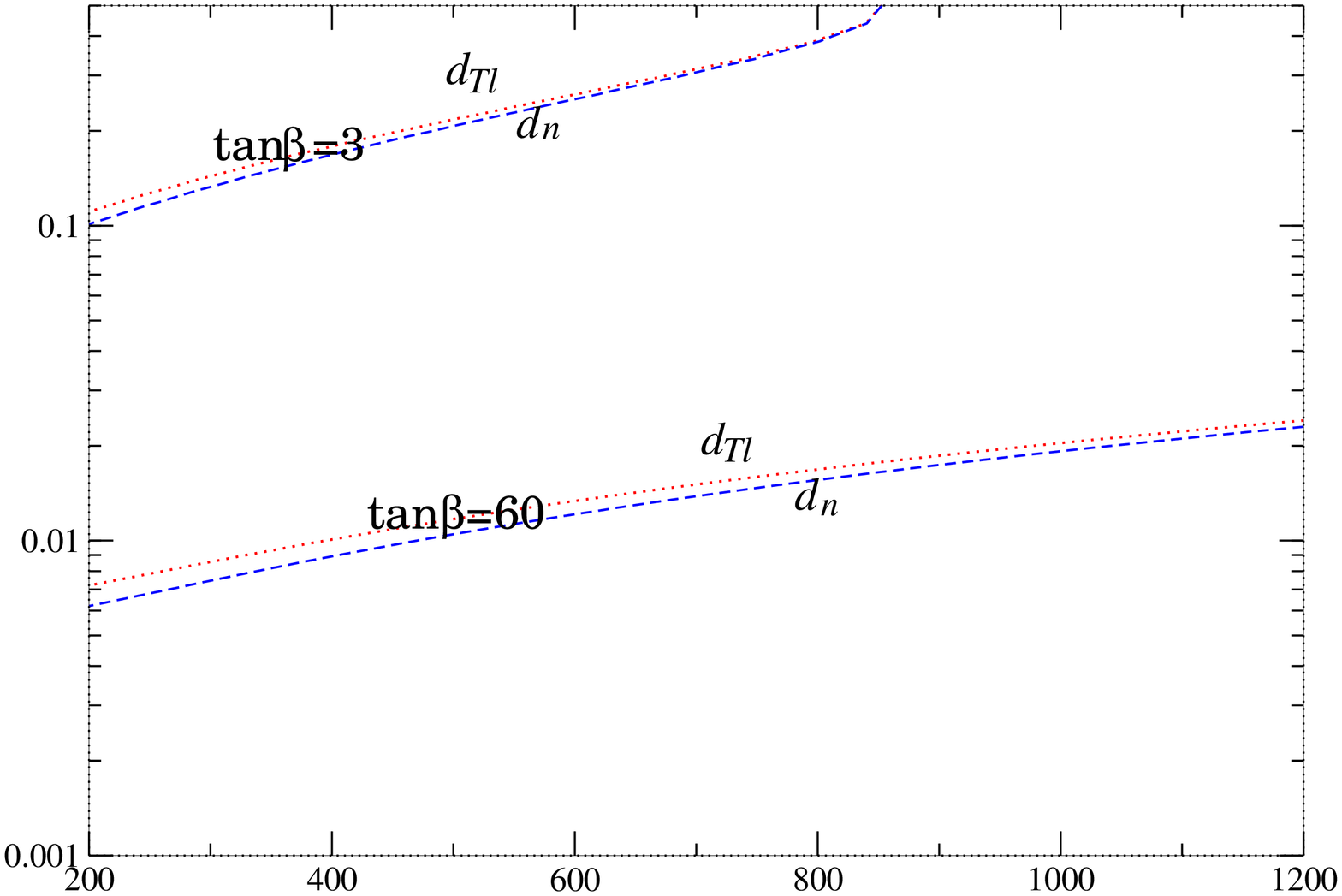,width=6.75cm,angle=-90}
\epsfig{file=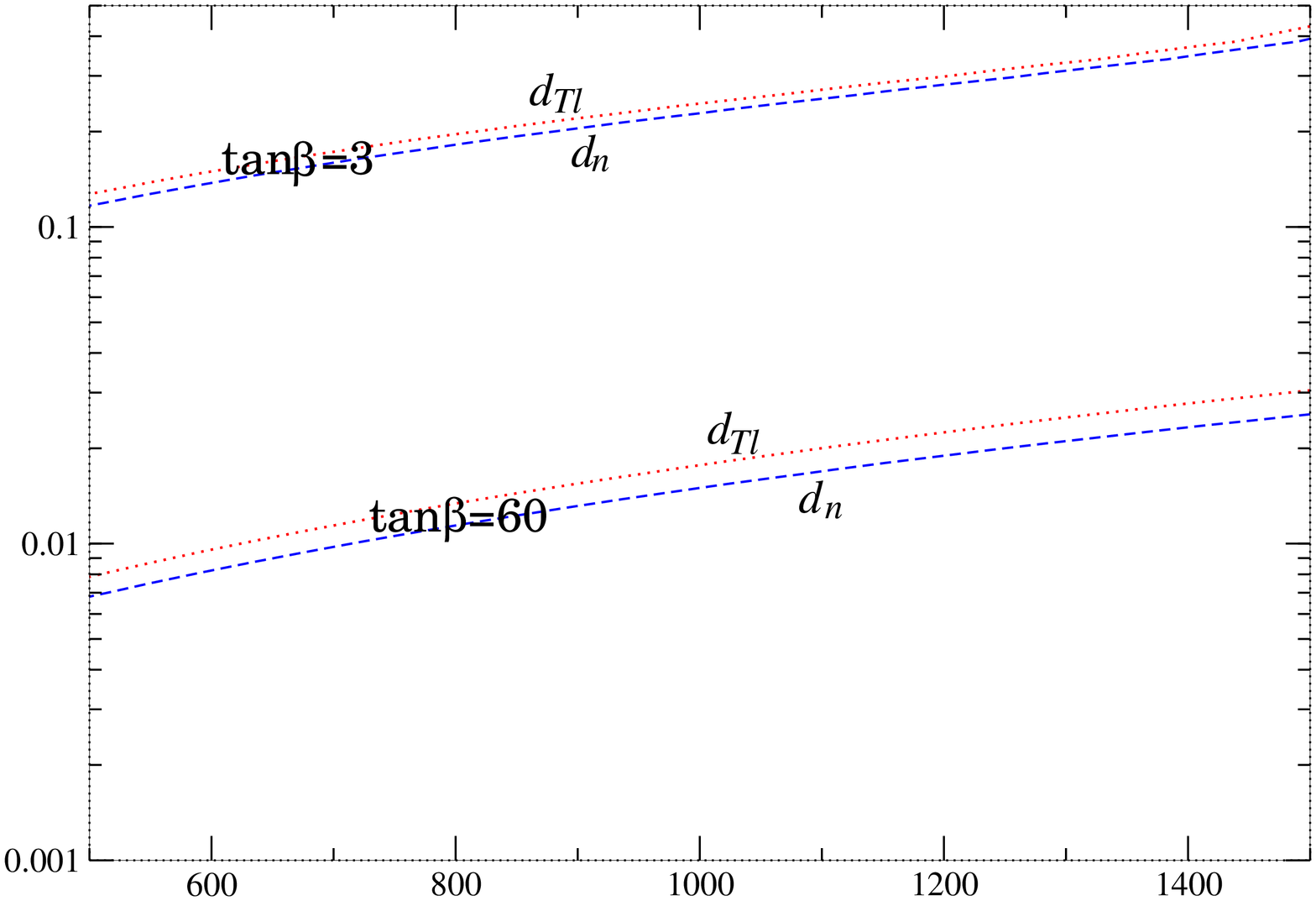,width=6.75cm,angle=-90}
\put(-400,-185){$M_2$ [GeV]} \put(-140,-185){$M_{H^{\pm}}$ [GeV]}
\put(-500,-40){$\phi_2(\pi)$} \put(-248,-40){$\phi_2(\pi)$} }
\caption{The $M_2$- and $M_{H^{\pm}}$- dependent constraints on
$\phi_2$ from neutron and Thallium EDMs in the case of heavy
first-two generations of sfermions ({\em case II}) as in Eq.
\ref{eq:case2}.} \label{fig:phi2b}
\end{figure}

\begin{figure}
\centerline{\epsfig{file=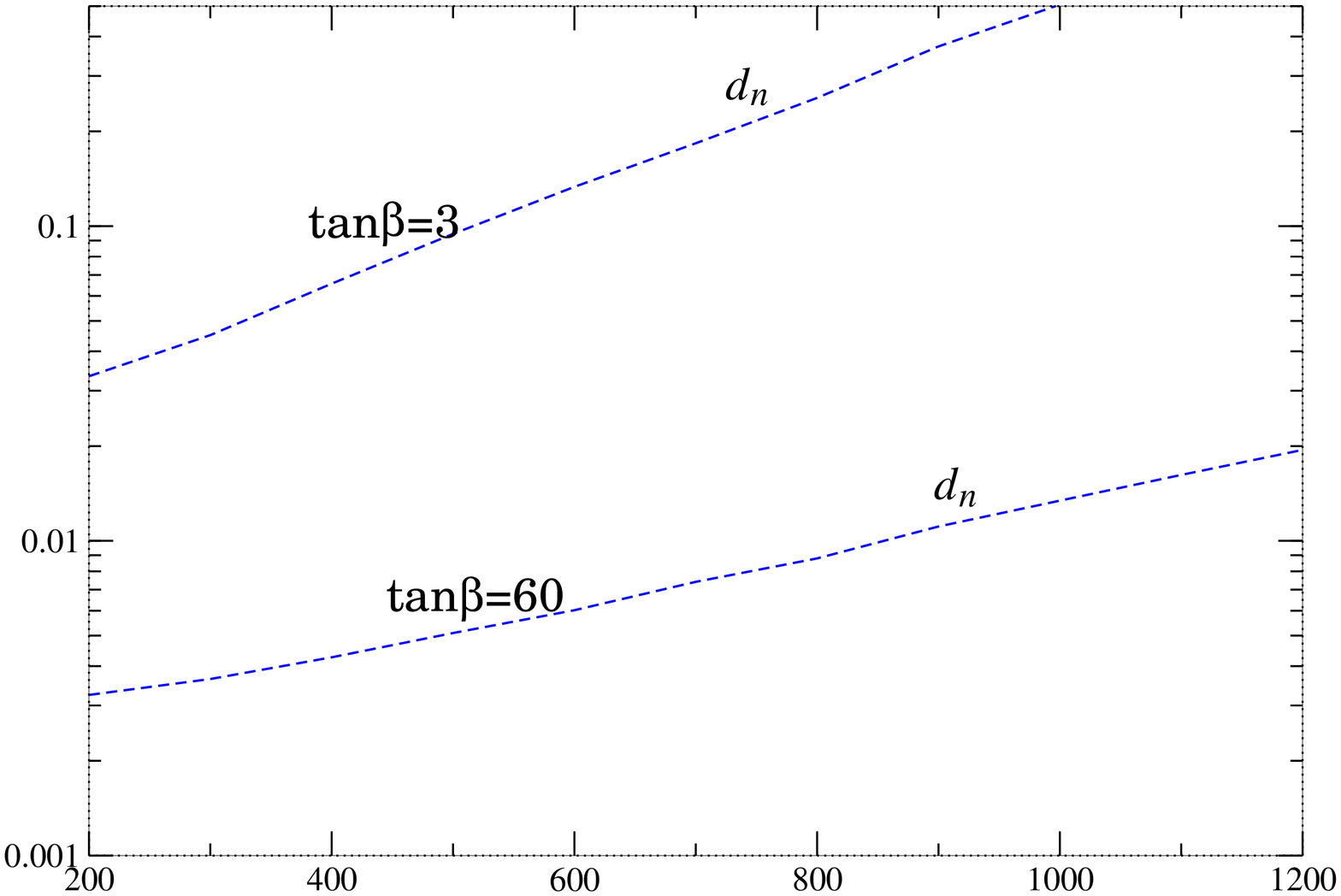,width=6.75cm,angle=-90}
\epsfig{file=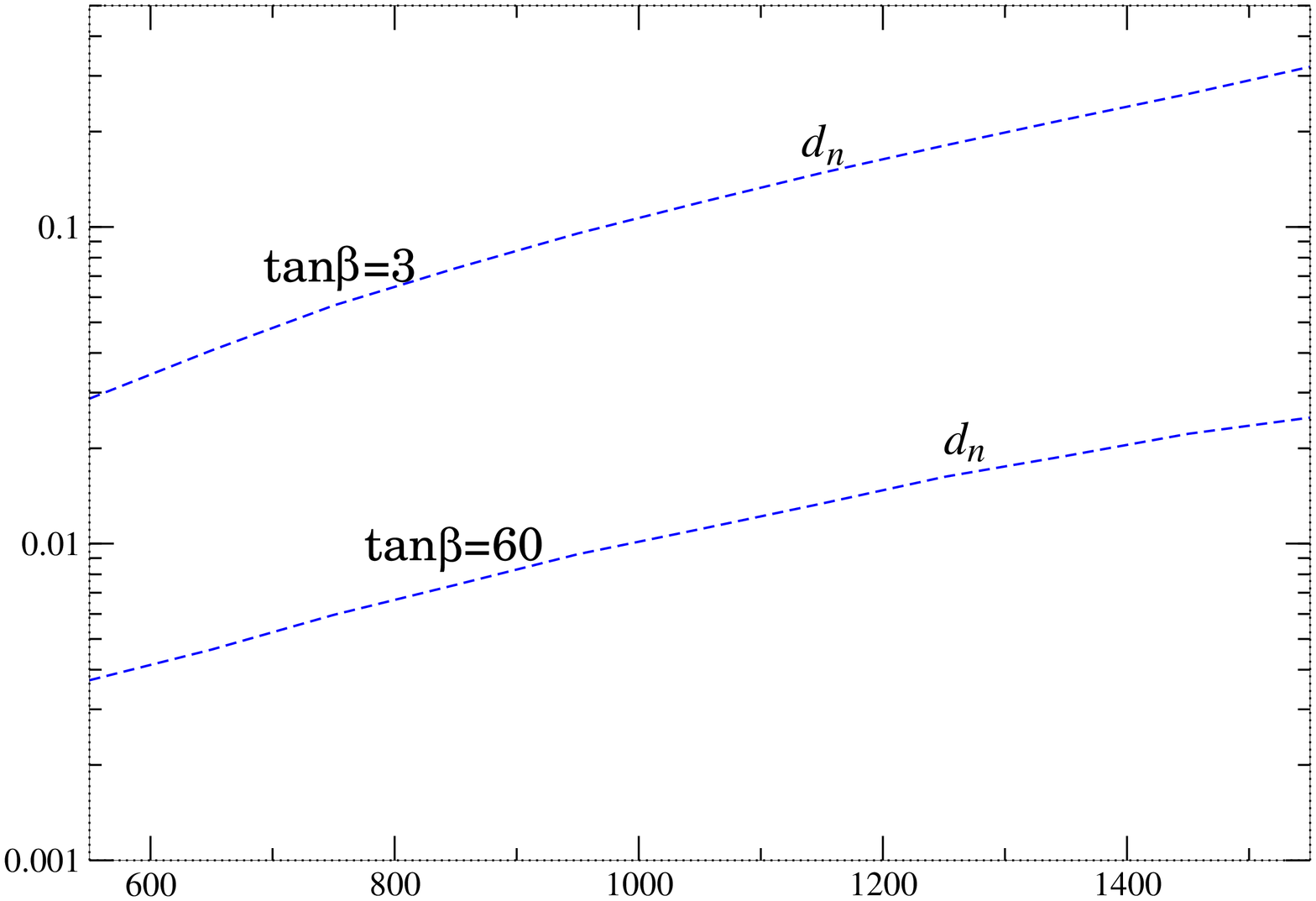,width=6.75cm,angle=-90}
\put(-410,-185){$(m_{L,R})_1$ [GeV]} \put(-140,-185){$M_3$ [GeV]}
\put(-500,-40){$\phi_3(\pi)$} \put(-248,-40){$\phi_3(\pi)$} }
\caption{The $(m_{L,R})_3$- and $M_3$- dependent constraints on
$\phi_3$ from neutron EDMs in the case of heavy first-two
generations of sfermions ({\em case II}) as in Eq. \ref{eq:case2}.}
\label{fig:phi3b}
\end{figure}

\section{Detailed analysis}
\label{sec:detail}

In the present study, we classify the MSSM CP-violating phases under
consideration into two groups,
\begin{itemize}
\item[(i)] phases in the higgsino-gaugino sector: $\phi_1$, $\phi_2$, and
$\phi_3$, and
\item[(ii)] phases in the sfermion sector:
$\phi_u$, $\phi_d$, $\phi_e$, $\phi_c$, $\phi_s$, $\phi_{\mu}$,
$\phi_t$, $\phi_b$, $\phi_{\tau}$,
\end{itemize}
and we study the two groups of phases individually.

\subsection{Phases in the higgsino-gaugino sector: $\phi_1$, $\phi_2$, and
$\phi_3$}

These three phases contribute to the EDM and Chromo-EDM operators both
at the one- and two-loop level; we therefore discuss here the constraints on
them for both cases: (I) with light and (II) with heavy first-two sfermion generations.

\vskip 0.2in

\noindent{\underline{Light sfermions}}

\vskip 0.2in

In the case where the first-two generation sfermions are light and
one-loop EDMs and Chromo-EDMs are not suppressed, the phases
$\phi_1$ and $\phi_2$ induce contributions to  $d_e$, while $\phi_{1,2,3}$ generate
($d_{u,d}$, $\tilde{d}_{u,d}$), and the four-fermion CP-odd operators
$C_{ff'}$. In addition,  $\phi_3$ also
induces a non-zero contribution to the 3-gluon operator $d^{3G}$ at two-loop order \cite{Dai:1990xh}

We find that as far as the constraints on $\phi_1$ are concerned,
the Mercury EDM, with its newest experimental bound, puts much more
stringent bounds -- namely by a factor of 10 or more on most of the parameter
space we consider -- than those from the neutron and Thallium EDM
limits. We illustrate the resulting limits in Figure \ref{fig:caseI}, top panels. For the phase $\phi_2$, instead, the Mercury EDM puts less
stringent limits, by a factor of a few, than the current Thallium
and neutron EDM bounds, illustrated in Figure \ref{fig:caseI}, middle panels. We find that this is due to the cancellation of
$\phi_2$ contributions to the electron EDM and quark chromo-EDMs that
generate Mercury EDM. In fact, in a different mass region where the cancelation
is not significant, the Mercury EDM constraint on $\phi_2$ may be
comparable or even stronger than the current Thallium and neutron
EDM bounds. For $\phi_3$, the new Mercury EDM bound puts a stronger
constraint than the neutron EDM bound, while the current Thallium EDM
bound is not stringent enough to put any constraint on $\phi_3$, due
to suppressed contributions from $C_{ff'}$ (see Figure \ref{fig:caseI}, lower panels).

Our results as a function of the relevant mass scales are summarized in Figure \ref{fig:caseI}.  In the upper penals, we show how the Mercury EDM constraint on
$\phi_1$ depends on ${\rm tan}\beta$ (set to 3 and to 60 in the upper and lower curves, respectively) and on the relevant mass scales
$(m_{L,R})_1$ (left panel) and $M_1$ (right panel). The most important impact of
$\phi_1$ on the Mercury EDM is through the neutralino one-loop
contribution to the quark Chromo-EDM $\tilde{d}^{\chi^0}_{u,d}$, where
the external gluon is only attached to squarks in the loop. In this
case, the dependence on the sfermion and gaugino masses are somewhat non-trivial: The allowed values for $\phi_1$ grow quickly with increasing
$(m_{L,R})_1$ (left panel), but rather slowly with increasing 
$M_1$ (right panel). As $(m_{L,R})_1$ increases to $1.2$ TeV,  $\phi_1$ can be as large as
$\pi/2(0.03\pi)$  for ${\rm
tan}\beta=3(60)$, indicating that a larger value for $\phi_1$ is allowed for smaller
${\rm tan}\beta$ when the one-loop contributions dominate.

In the two middle panels of Fig. \ref{fig:caseI}, we show how the neutron (blue dashed lines) and the Thallium (red dotted lines) EDM
constraints on $\phi_2$ depend on ${\rm tan}\beta$ and on the relevant mass
scales $(m_{L,R})_1$ and $M_2$. We notice that the current neutron and Thallium EDM
bounds put comparable constraints on the wino phase $\phi_2$. The most important
contributions from $\phi_2$ to the neutron and Thallium EDMs are through
the chargino one-loop contribution to quark and electron EDMs
$d^{\chi^{\pm}}_{u,d,e}$, where the external photon is attached to
both sfermions and charginos in the loop. In this case, the
dependence on the sfermion and gaugino masses resemble the case
of the neutralino loop. A comparison of the left and right panels shows that the allowed values of $\phi_2$ grows a little
faster with increasing $(m_{L,R})_1$ than with increasing $M_2$. For ${\rm
tan}\beta=3(60)$, the phase $\phi_2$ is allowed to be within
$10^{-2} \pi(0.6\times10^{-3}\pi)$ for $(m_{L,R})_1$ up to $1.2$
TeV, and $0.5\times10^{-2}\pi$($0.3\times10^{-3}\pi$) for $M_2$ up
to $1.25$ TeV.

In the lower panels of Fig. \ref{fig:caseI}, we finally show how the constraints on $\phi_3$
depend on ${\rm tan}\beta$ and relevant mass scales $(m_{L,R})_1$ (left)
and $M_3$ (right). The black line indicates the Mercury EDM constraint, while the blue dashed line refers to the neutron EDM. We observe that the constraints on $\phi_3$ do not
monotonically increase with mass scales, due to non-trivial cancelations among
$\tilde{d}_{u,d}$, $d_{u,d}$, and $d^{3G}$ in their contributions to
the neutron EDM, and to cancellations among different contributions to
$\tilde{d}_{u,d}$, which dominate the Mercury EDM. However, there is
no common region where both the neutron and the Mercury EDM constraints are
suppressed; hence, the cancellations among various $\phi_3$-dependent contributions never open a region in this portion of parameter space region where this CP-violating phase can be large, independent of the values of the other phases. For ${\rm tan}\beta=3(60)$, the phase $\phi_3$ is allowed
to be within $10^{-2}\pi(0.5\times10^{-3}\pi)$ for $(m_{L,R})_1$ up
to $1.2$ TeV, and $10^{-2}\pi$($0.2\times10^{-3}\pi$) for $M_3$ up
to $1.55$ TeV.

\vskip 0.2in

\newpage
\noindent{\underline{Heavy sfermions}}

\vskip 0.2in

In the case where the first-two generations of sfermions are heavy
(case II), and therefore the one-loop EDMs and Chromo-EDMs are
suppressed, the gaugino phases $\phi_1$ and $\phi_2$
%, and $\phi_3$ 
still induce a non-vanishing 4-fermion CP-odd operator $C_{ff'}$ due to loop-induced mixing between the CP-even and CP-odd Higgses that are exchanged between the fermions.  In addition,
$\phi_1$ and $\phi_2$ induce $d_e$ and
$d_{u,d}$ at the two-loop level, while $\phi_3$ induces 3-gluon
operator $d^{3G}$ and Chromo-EDM $\tilde{d}_{u,d}$ at the two-loop
level.

Fig.~\ref{fig:phi1b} shows that in case II (heavy sfermions) the
phase $\phi_1$ is not constrained by current EDM bounds at all. This
has important consequences for scenarios where the baryon asymmetry
in the universe is generated via the mechanism of electroweak
baryogenesis: a non-vanishing and large enough $\phi_1$ can generate
the observed baryon asymmetry via the bino-driven scenario in the
context of electroweak baryogenesis, even if $\phi_2=0$
\cite{Li:2008ez}. In Fig.~\ref{fig:phi1b}, where we set $\phi_2=0$,
we show contours of constant values for the neutron and Thallium
EDMs that are below the current bounds,
at $\tan\beta=3$ and 60. Notice that, in contrast to the situation for $\tan\beta=60$, at
$\tan\beta=3$ cancellations occur between the $WW$ and the $W^\pm
H^\mp$ contributions, the only non-vanishing graphs for $\phi_2=0$
\cite{Li:2008kz}. In particular, Ref.~ \cite{Li:2008kz} pointed out
that these two contributions have opposite signs, and if $\tan\beta$
has a value such that the two are comparable, then cancellations are
possible, and indeed they occur for $M_1\sim320$ GeV and for
$m_{H}\sim420$ GeV in our setup, as shown in Fig.~\ref{fig:phi1b}.
Similar cancellations do not take place at large $\tan\beta$, where
the $W^\pm H^\mp$ contribution dominates.

If instead the wino phase $\phi_2$ is non-vanishing
(Fig.~\ref{fig:phi2b}), two-loop contributions are much larger (see
Ref.~ \cite{Li:2008kz} for a discussion on what makes wino-driven
two-loop EDMs typically a factor of 50-100 larger than bino-driven
ones). Specifically, with our choice of parameters, the limits on
$\phi_2$ as a function of $M_2$ from the EDM bounds of Thallium atom
and neutron range from 0.006 $\pi$ for small $M_2=200$ GeV up to
0.02 $\pi$ for $M_2\sim1$ TeV. A similar dependence is found for the
second mass scale entering the two-loop contribution, namely the
heavy Higgs sector, where we find that $\phi_2$ must be smaller than
0.007 $\pi$ for small $m_{H^\pm}\sim 500$ GeV and than 0.03 $\pi$
for $m_{H^\pm}\sim 1500$ GeV.

In case II, the phase $\phi_3$ induces EDMs of neutron, Thallium,
and Mercury. Since it does not induce an electron EDM $d_e$, its
contribution to Thallium EDM is highly suppressed. The phase $\phi_3$ could induce
sizable Mercury EDM through generating $\tilde{d}_{u,d}$ at
two-loop. However, the most stringent constraint comes from the current
neutron EDM bound. In Fig. \ref{fig:phi3b}, we show how the
constraint on $\phi_3$ from neutron EDM bound depends on ${\rm
tan}\beta$ and relevant mass scales $(m_{L,R})_3$ and $M_3$. We
observe that for ${\rm tan}\beta=3(60)$, the $\phi_3$ is allowed to
be as large as $\pi/2(0.02\pi)$ for $(m_{L,R})_1$ as heavy as $1.2$ TeV, and
$0.3\times\pi(0.02\pi)$ for $M_3\leq 1.55$ TeV.

\begin{figure}
\centerline{\epsfig{file=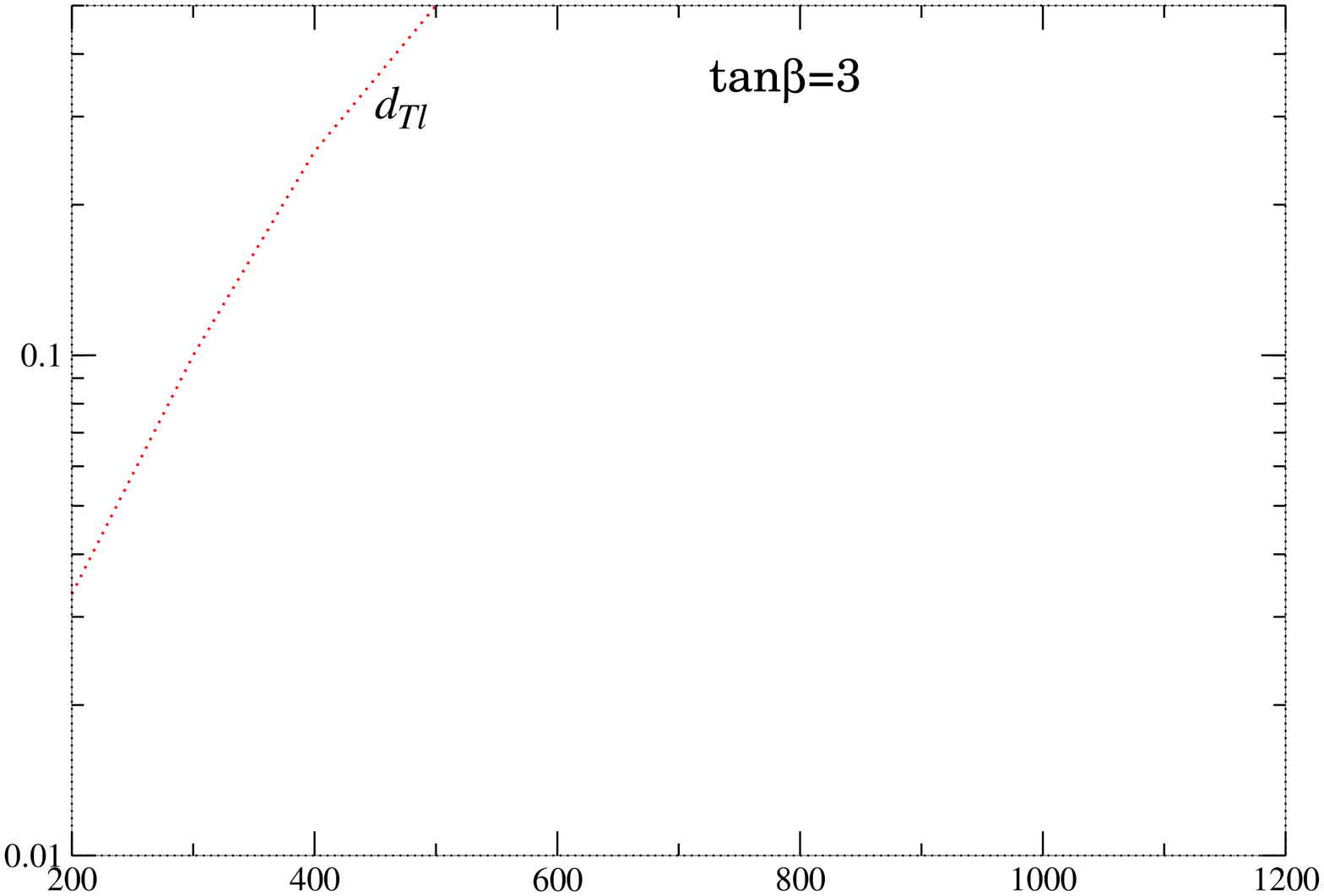,width=6.75cm,angle=-90}
\epsfig{file=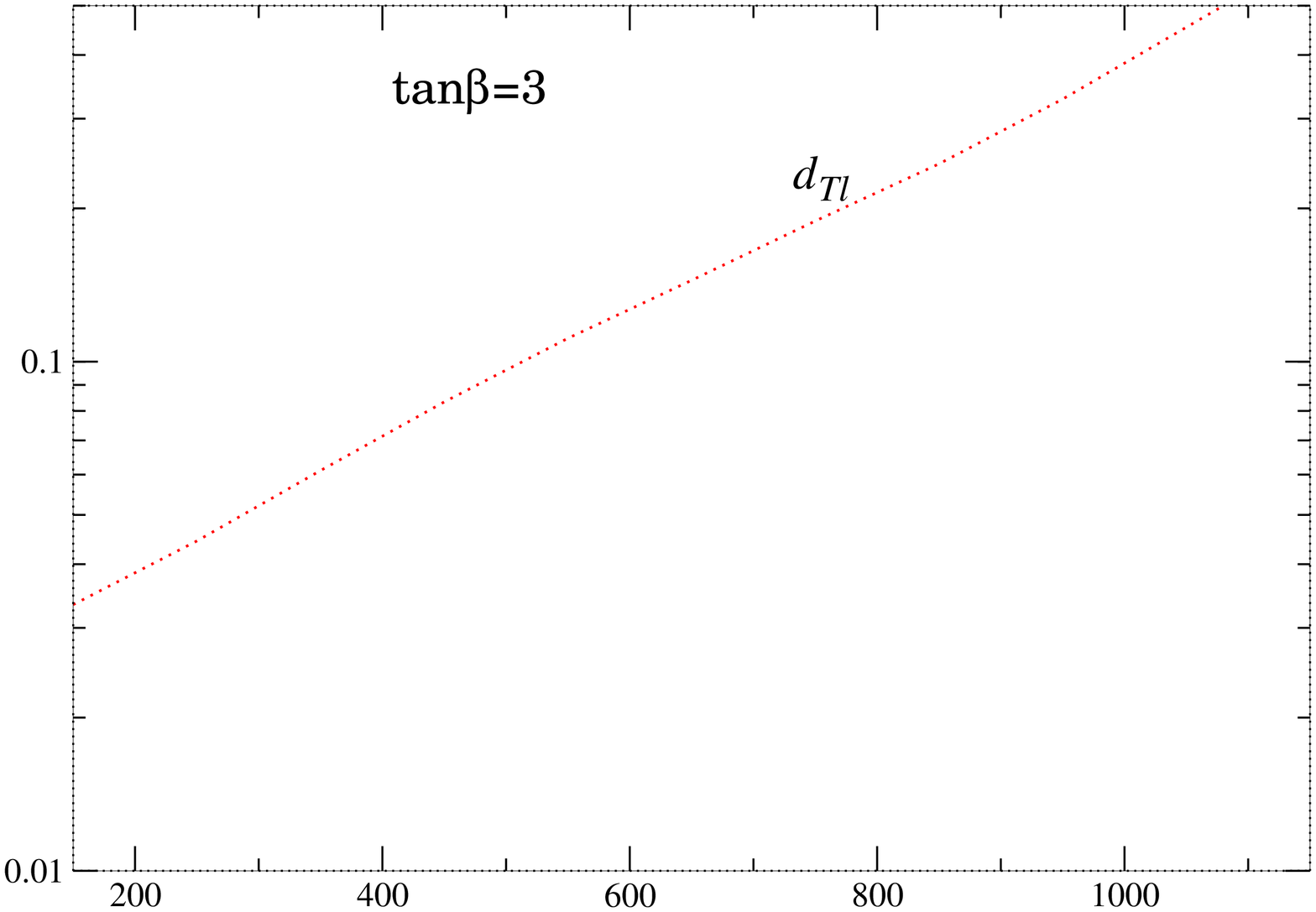,width=6.75cm,angle=-90}
\put(-410,-185){$(m_{L,R})_1$ [GeV]} \put(-140,-185){$M_1$ [GeV]}
\put(-500,-40){$\phi_e(\pi)$} \put(-248,-40){$\phi_e(\pi)$} }
\caption{The $(m_{L,R})_1$- and $M_1$- dependent constraints on
$\phi_e$ from Thallium EDM in the case with light first-two
generations of sfermions as in Eq. \ref{eq:case1}. The ${\rm
tan}\beta=3$ is used in making the plot. The constraints
corresponding to other values of ${\rm tan}\beta$ are not shown, as
the ${\rm tan}\beta$ dependence is found to be rather weak.}
\label{fig:phie}
\end{figure}

\begin{figure}
\centerline{\epsfig{file=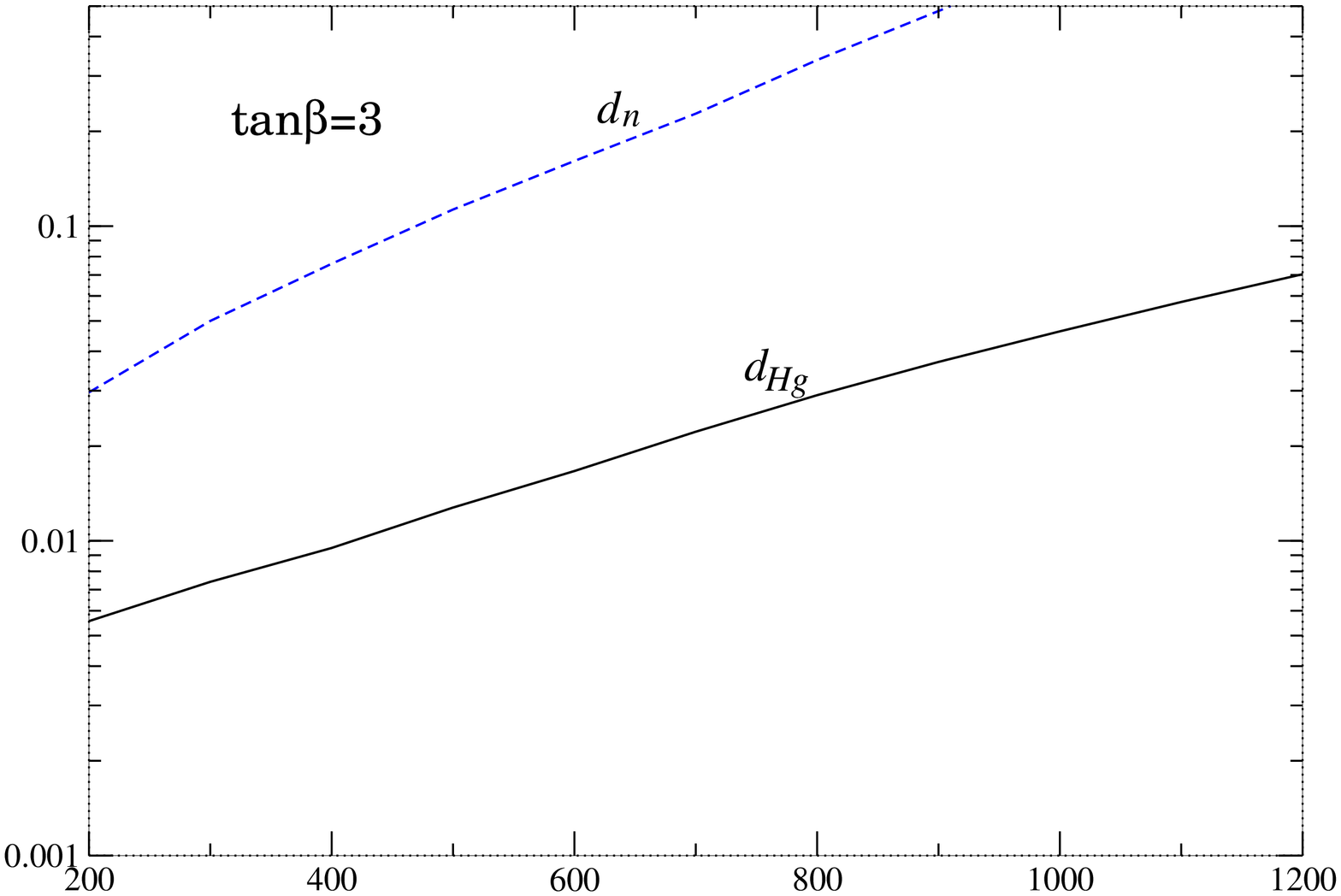,width=6.75cm,angle=-90}
\epsfig{file=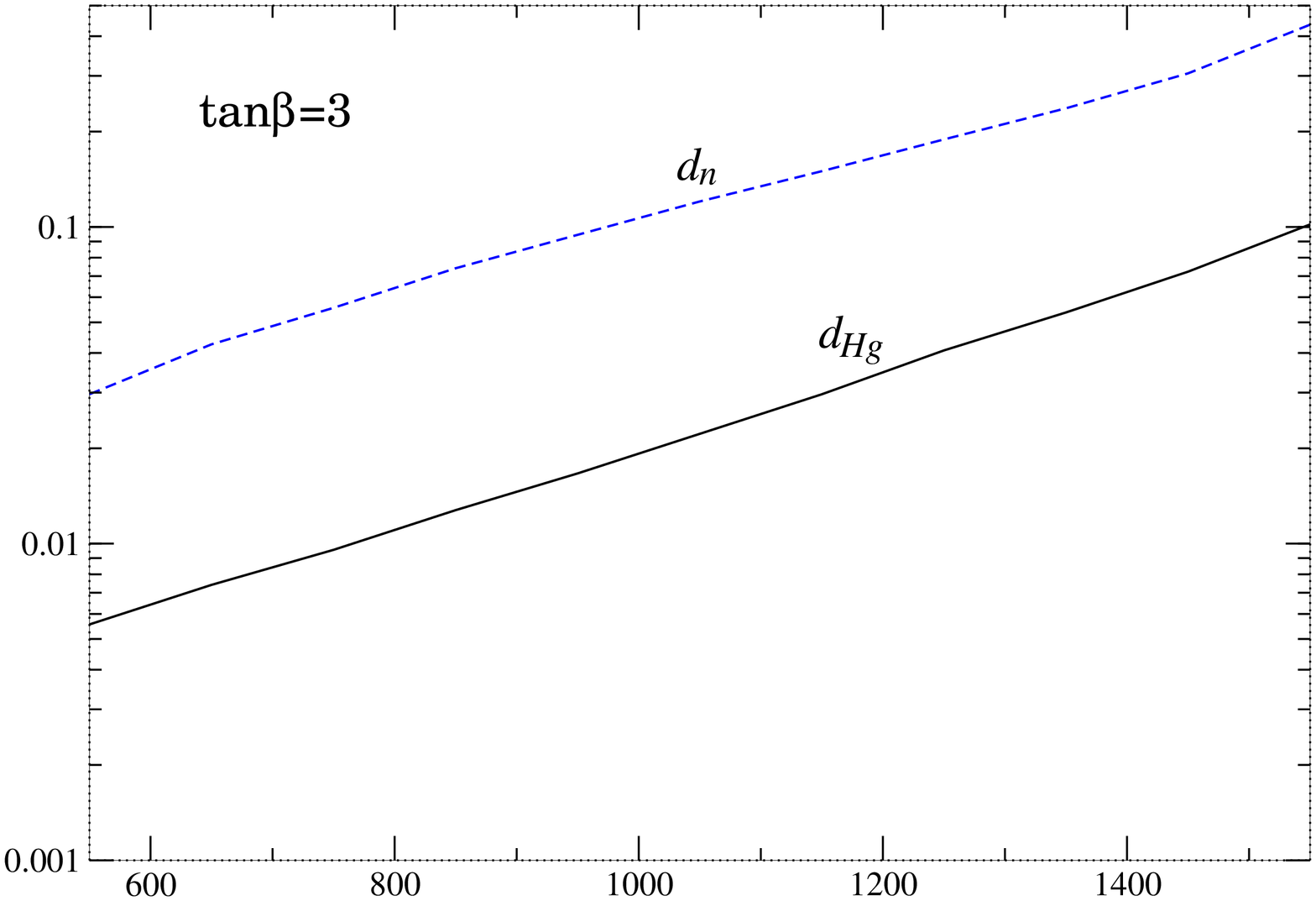,width=6.75cm,angle=-90}
\put(-410,-185){$(m_{L,R})_1$ [GeV]} \put(-140,-185){$M_3$ [GeV]}
\put(-500,-40){$\phi_u(\pi)$} \put(-248,-40){$\phi_u(\pi)$} }
\caption{The $(m_{L,R})_1$- and $M_3$- dependent constraints on
$\phi_u$ from neutron and Mercury EDMs in the case with light
first-two generations of sfermions as in Eq. \ref{eq:case1}. The
${\rm tan}\beta=3$ is used in making the plot. The constraints
corresponding to other values of ${\rm tan}\beta$ are not shown, as
the ${\rm tan}\beta$ dependence is found to be rather weak.}
\label{fig:phiu}
\end{figure}

\begin{figure}
\centerline{\epsfig{file=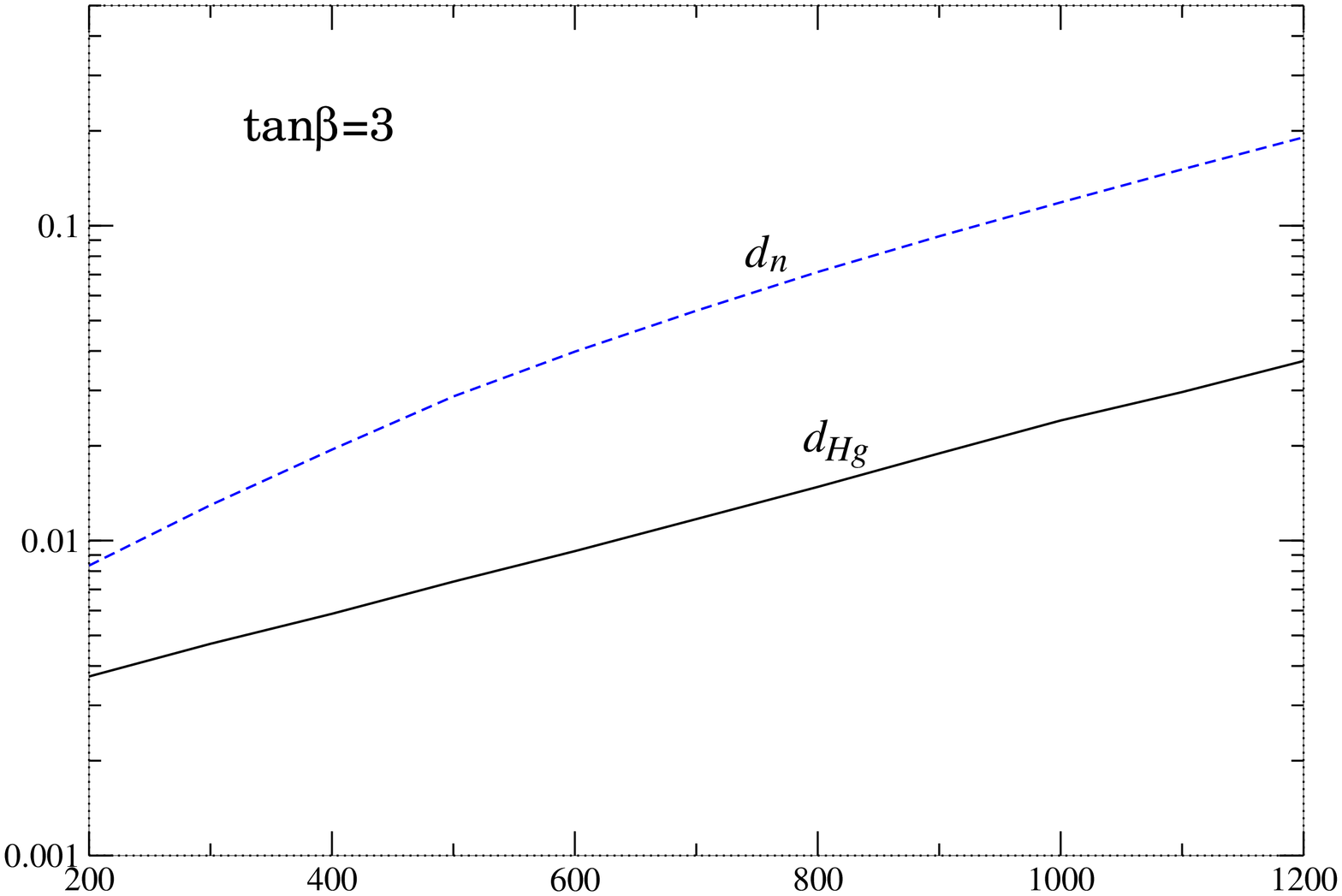,width=6.75cm,angle=-90}
\epsfig{file=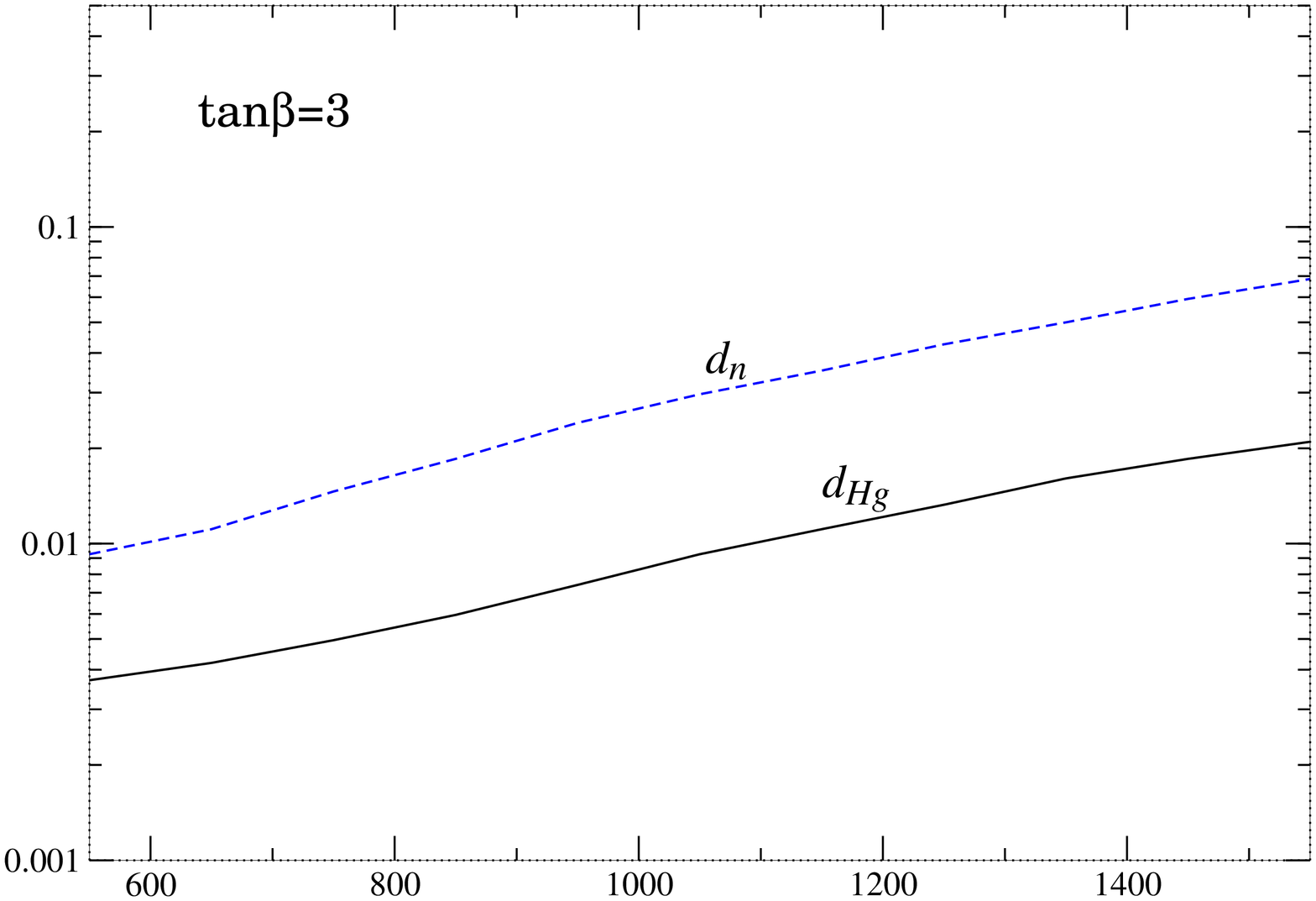,width=6.75cm,angle=-90}
\put(-410,-185){$(m_{L,R})_1$ [GeV]} \put(-140,-185){$M_3$ [GeV]}
\put(-500,-40){$\phi_d(\pi)$} \put(-248,-40){$\phi_d(\pi)$} }
\caption{The $(m_{L,R})_1$- and $M_3$- dependent constraints on
$\phi_d$ from neutron and Mercury EDMs in the case with light
first-two generations of sfermions as in Eq. \ref{eq:case1}. The
${\rm tan}\beta=3$ is used in making the plot. The constraints
corresponding to other values of ${\rm tan}\beta$ are not shown, as
the ${\rm tan}\beta$ dependence is found to be rather weak.}
\label{fig:phid}
\end{figure}

\begin{figure}
\centerline{\epsfig{file=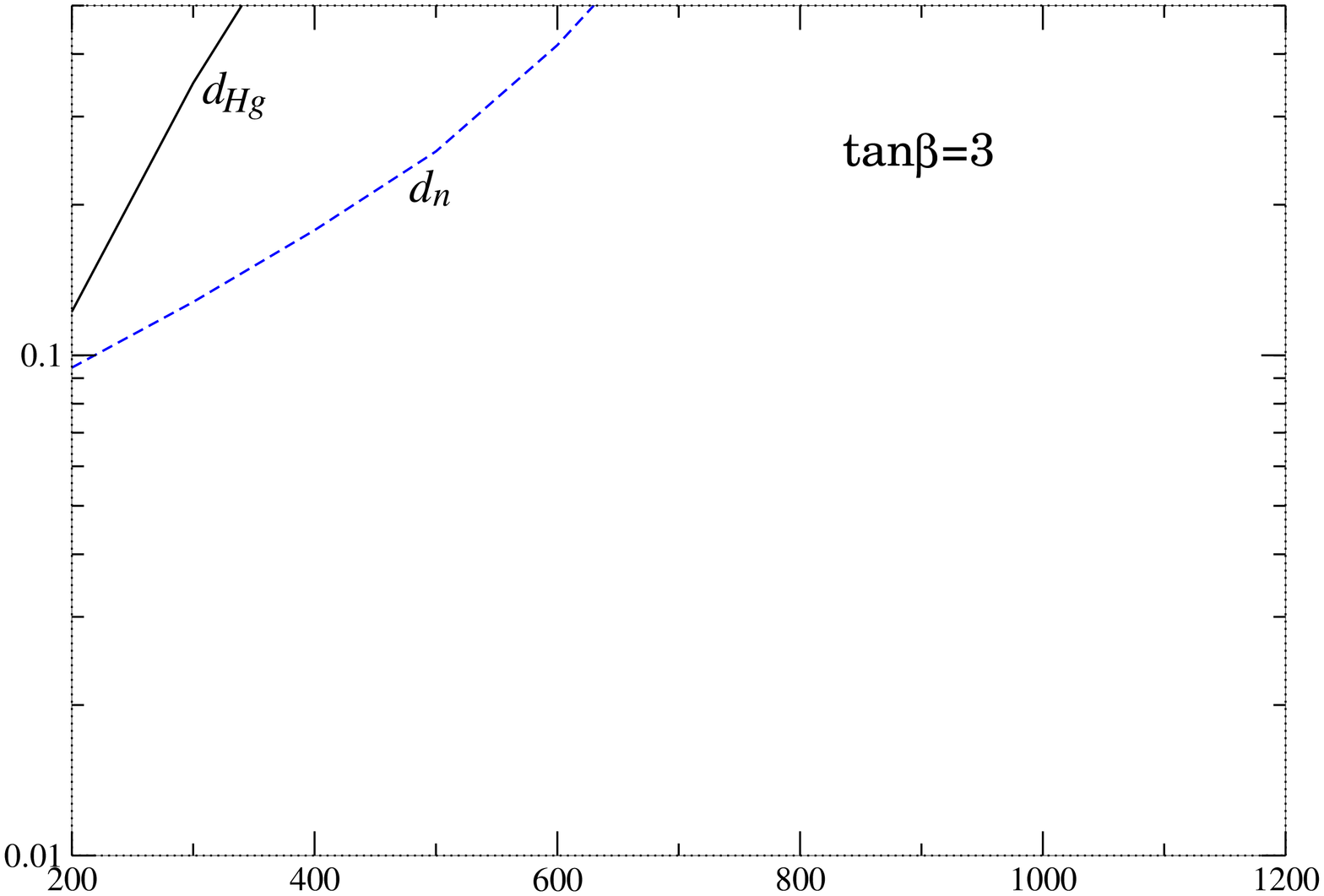,width=6.75cm,angle=-90}
\epsfig{file=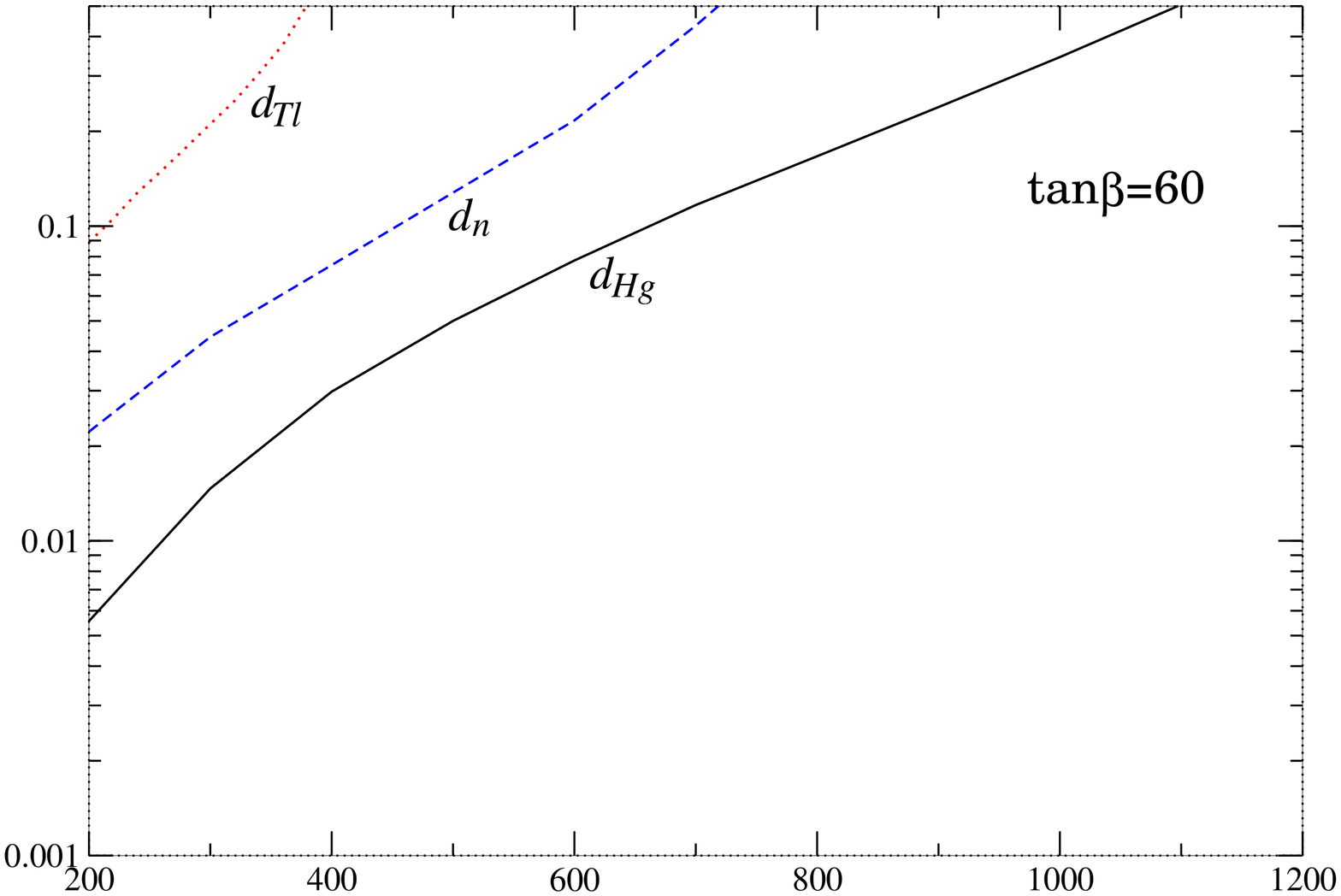,width=6.75cm,angle=-90}
\put(-410,-185){$(m_{L,R})_3$ [GeV]} \put(-160,-185){$(m_{L,R})_3$
[GeV]} \put(-500,-40){$\phi_t(\pi)$} \put(-248,-40){$\phi_t(\pi)$} }
\caption{The $(m_{L,R})_3$-dependent constraints on $\phi_t$ from
neutron, Thallium, and Mercury EDMs.} \label{fig:phit}
\end{figure}

\begin{figure}
\centerline{\epsfig{file=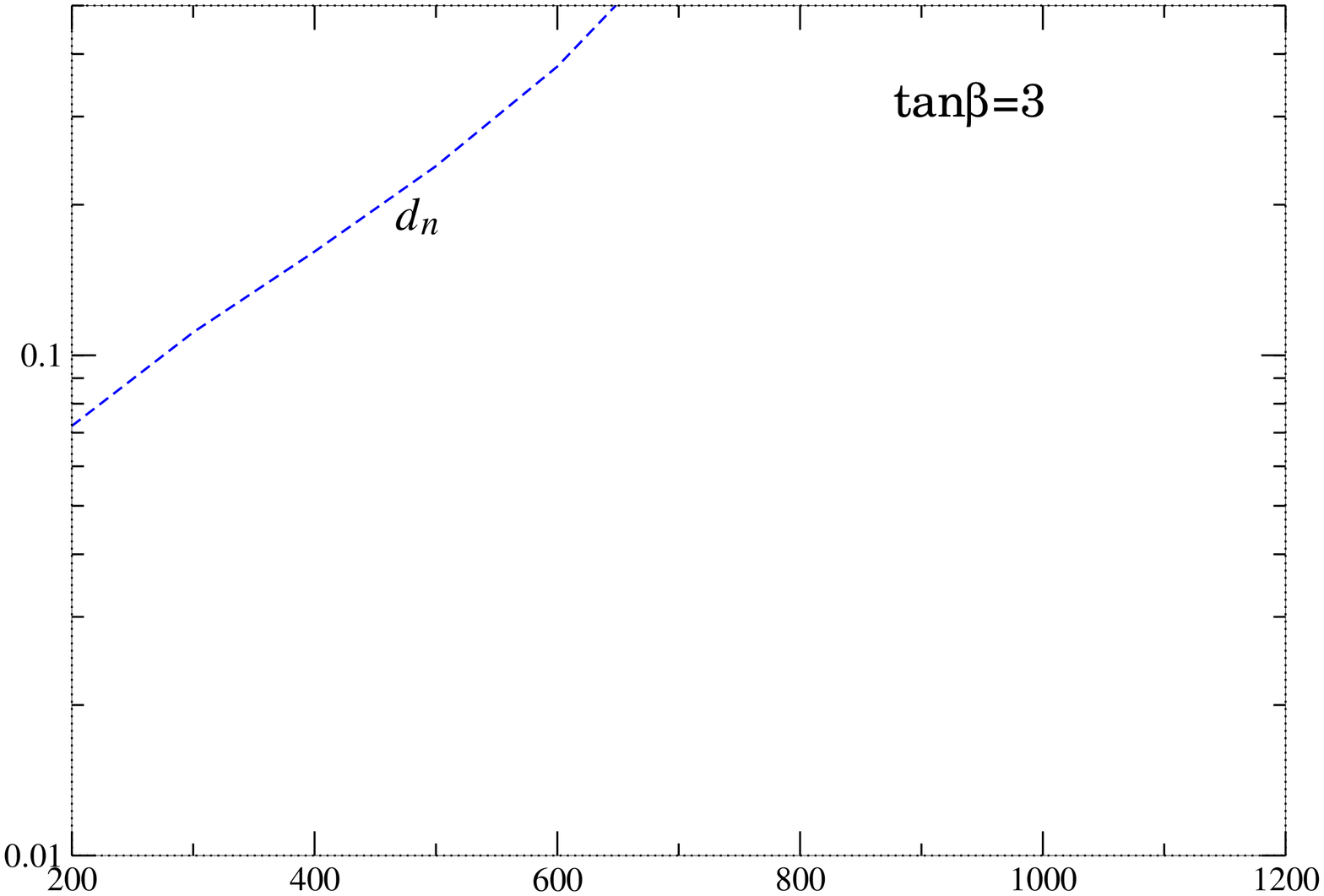,width=6.75cm,angle=-90}
\epsfig{file=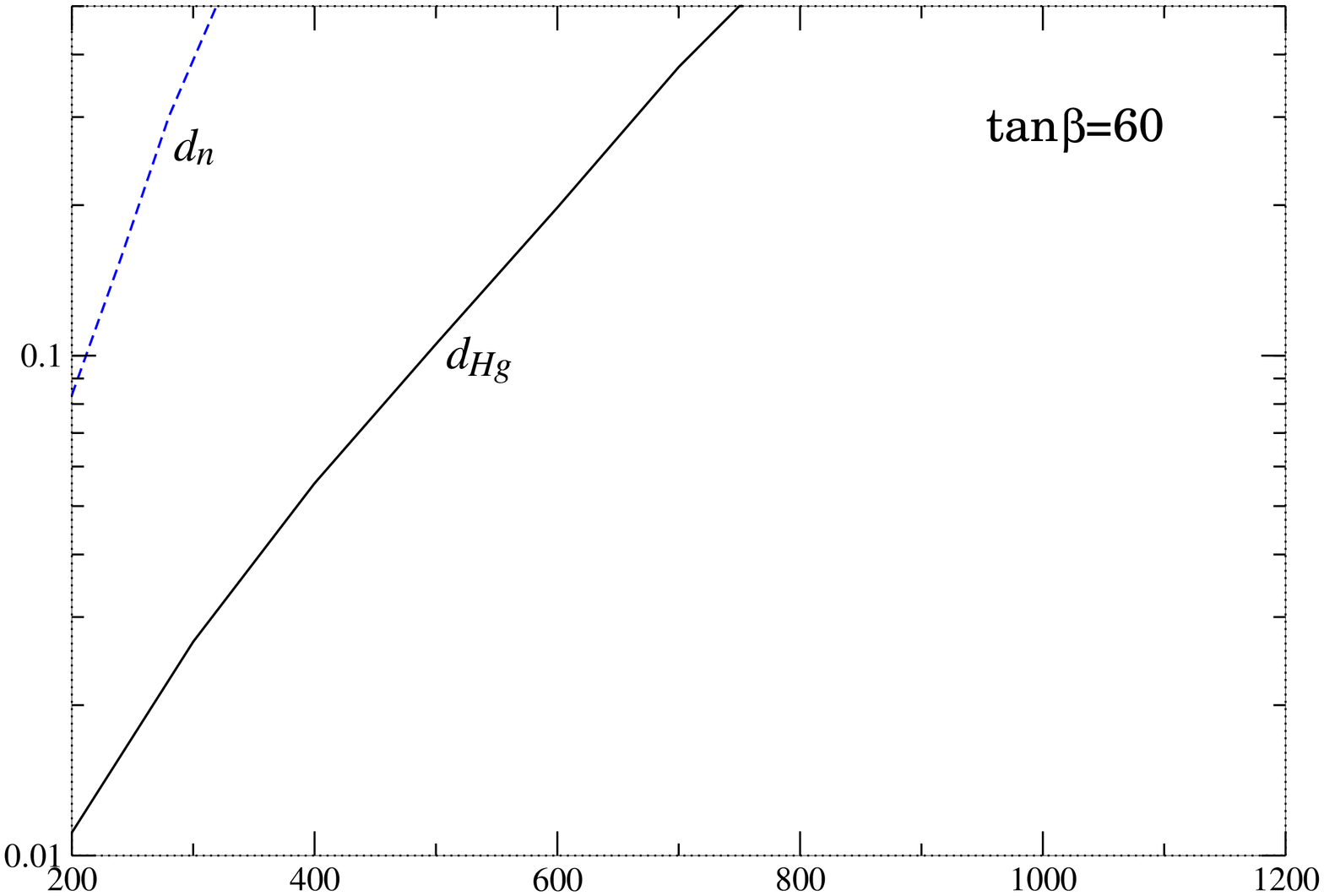,width=6.75cm,angle=-90}
\put(-410,-185){$(m_{L,R})_3$ [GeV]} \put(-160,-185){$(m_{L,R})_3$
[GeV]} \put(-500,-40){$\phi_b(\pi)$} \put(-248,-40){$\phi_b(\pi)$} }
\caption{The $(m_{L,R})_3$-dependent constraints on $\phi_b$ from
neutron and Mercury EDMs.} \label{fig:phi1abc}
\end{figure}

\subsection{Phases in the Sfermion sector: $\phi_e$, $\phi_u$, $\phi_d$,
$\phi_{\mu}$, $\phi_c$, $\phi_s$, $\phi_t$, $\phi_b$, $\phi_{\tau}$}

\begin{table}[!b]
\caption{Summary of how the CP-violating phases in MSSM are
constrained by current EDM bounds of neutron, Thallium, and Mercury
atom, for both case I (with light first-two generations of sfermions
as in Eq. \ref{eq:case1}) and case II (with heavy first-two
generations of sfermions as in Eq. \ref{eq:case2}). We refer as
"weakly constrained" if the phase can reach $\pi/2$ with relevant
mass scales within about 1 TeV, and "strongly constrained"
otherwise.}
\begin{center}
\label{tab:phase2edm}
\begin{tabular}{|c|c|c|c|c|c|c|}
\hline &
\multicolumn{3}{|c|}{CASE I} & \multicolumn{3}{|c|}{CASE II} \\
\hline phases &  $d_{\rm Tl}$ & $d_n$ & $d_{\rm Hg}$ & $d_{\rm Tl}$
&
$d_n$ & $d_{\rm Hg}$ \\
\hline $\phi_1$ & weakly & weakly & weakly w. small ${\rm tan}\beta$ & not & not & not \\
\hline $\phi_2$ & strongly & strongly & strongly & weakly w. small ${\rm tan}\beta$ & weakly w. small ${\rm tan}\beta$ & not \\
\hline $\phi_3$ & not & strongly & strongly & not & weakly w. small ${\rm tan}\beta$ & weakly \\

\hline $\phi_{e}$ & weakly & not & not & not & not & not \\
\hline $\phi_{u}$ & not & weakly & strongly & not & not & not \\
\hline $\phi_{d}$ & not & strongly & strongly & not & not & not \\

\hline $\phi_{\mu}$ & not & not & not & not & not & not \\
\hline $\phi_{c}$ & not & not & not & not & not & not \\
\hline $\phi_{s}$ & not & not & not & not & not & not \\

\hline $\phi_{t}$ & weakly & weakly & weakly & weakly & weakly & weakly \\
\hline $\phi_{b}$ & not & weakly & weakly & not & weakly & weakly \\
\hline $\phi_{\tau}$ & not & not & not & not & not & not \\
\hline
\end{tabular}
\end{center}
\end{table}

We discuss here how CP violating phases in the sfermion sector are constrained generation by generation.
The phases associated with the first sfermion generation, namely $\phi_e$, $\phi_u$ and
$\phi_d$, yield significant contributions to the electron and u- and d-quark EDMs, as well as to the Chromo-EDM,
all only at the one-loop level. Thus, we only discuss here the case where the
first-two generations of sfermions are light (case I) as in the case where sfermions are heavy all of these contributions asymptotically vanish.

Besides contributions to the 4-fermion CP-odd operators $C_{ff'}$ which
are in general small, the phase $\phi_e$ contributes to the electron
EDM, and is therefore constrained by the Thallium EDM. The phases
$\phi_u$ and $\phi_d$ contribute to the EDM and to the Chromo-EDM of
quarks, and are therefore constrained by both the neutron and the
Mercury EDMs. The relevant mass scales in these one-loop
contributions are $(m_{L,R})_1$ and $M_{1,2,3}$. However, since the
one-loop contribution involving neutralinos dominates  $d_e$, the constraint on $\phi_e$ depends mainly on
$(m_{L,R})_1$ and $M_{1}$. On the other hand, constraints on the
squark CP violating phases $\phi_u$ and $\phi_d$ mainly depend on $(m_{L,R})_1$ and $M_{3}$, since here the one-loop
contribution involving gluinos dominates 
$d_{u,d}$ and  $\tilde{d}_{u,d}$. The dependence on ${\rm
tan}\beta$ is found to be rather weak.

As shown in Fig. \ref{fig:phie}, the allowed value of $\phi_e$ grows
much faster with increasing $(m_{L,R})_1$ than $M_1$, due to the
dominant neutralino one-loop contribution. Constraints on this
phase are rather weak and they reach $\pi/2$ for
$m_{{L,R}_1}$ of only $500$ GeV, and for $M_1$ of $1.15$ TeV.

As far as constraints on $\phi_u$ and $\phi_d$ are concerned, the new
experimental limit on the Mercury EDM places a more stringent constraint (by a factor of
a few) than the current neutron EDM bound. As shown in Figs.
\ref{fig:phiu} and \ref{fig:phid}, which display the $(m_{L,R})_1$- and
$M_3$- dependent constraints on $\phi_u$ and $\phi_d$, respectively,
the phase $\phi_u$($\phi_d$) is constrained to be  $\leq 0.07
\pi$($0.04\pi$) for $(m_{L,R})_1\leq1.2$ TeV, and
$0.1\pi$($0.02\pi$) for $M_3\leq 1.55$ TeV.

The phases associated with the third generation sfermions  $\phi_t$, $\phi_b$,
and $\phi_{\tau}$ do not directly induce an electron or quark EDM, or a
Chromo-EDM at the one-loop level, and therefore there is no difference here
between the case with light and heavy first-two generations of
sfermions. The third generation CP violating squark phases
contribute to the electron and to the quark EDMs, 
Chromo-EDM s, 3-gluon operator, and 4-fermion
operators at the two-loop level, and they all induce contributions to the EDMs of the
neutron, Thallium and Mercury atoms. Our numerical study indicates
that the current Thallium EDM bound is in general weaker than
neutron and Mercury EDM bounds in constraining $\phi_t$, and does
not put any constraint on $\phi_b$. The relative strength of the
neutron and Mercury EDM bounds on $\phi_t$ and $\phi_b$ depends on
${\rm tan}\beta$, which drives the relative size of the Yukawa
couplings of the bottom and top quarks. As shown in Fig.
\ref{fig:phit}, at small ${\rm tan}\beta$, the neutron EDM constraint is
stronger than that of the Mercury EDM , while at large ${\rm tan}\beta$,
the Mercury EDM constraint becomes dominant. The $\phi_t$ and $\phi_b$ are
rather loosely bounded and can reach $\pi/2$ for $(m_{L,R})_3$ of a
few hundred GeV. We find that the $\phi_{\tau}$ phase is not constrained by current EDM
bounds at all.

Similarly, we find that the CP violating phases in the second generation of sfermions
$\phi_{\mu}$, $\phi_c$, and $\phi_s$ are essentially unconstrained by experimental EDM limits.

\subsection{Correlated constraints on the most strongly constrained phases $\phi_2$, $\phi_3$ and $\phi_{u,d}$}

\begin{figure}
\begin{center}
\epsfig{file=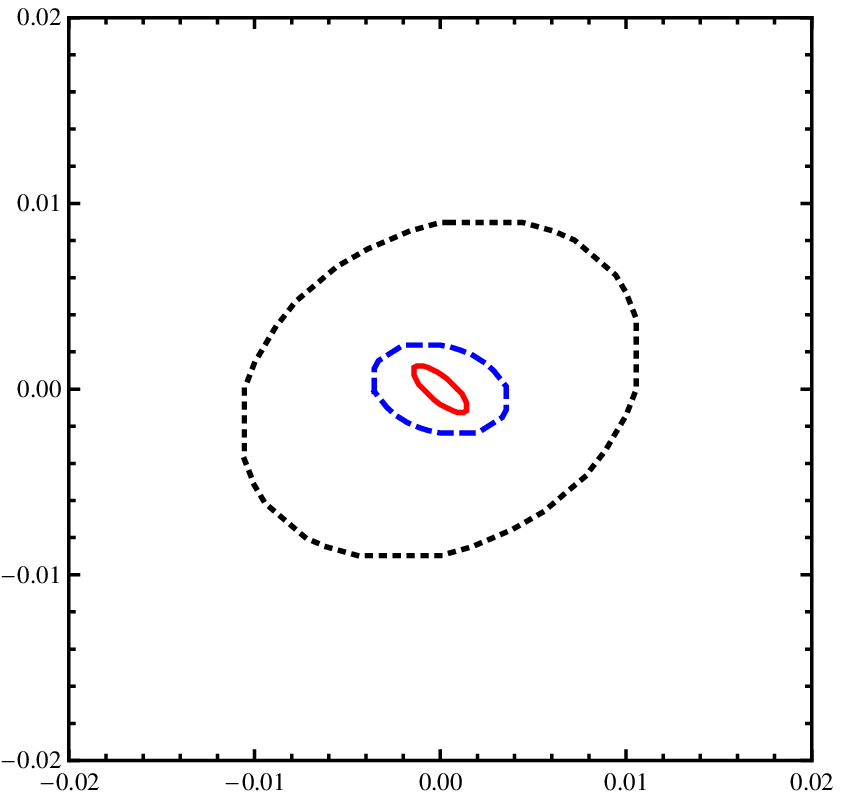,width=7.25cm}~~~~
\epsfig{file=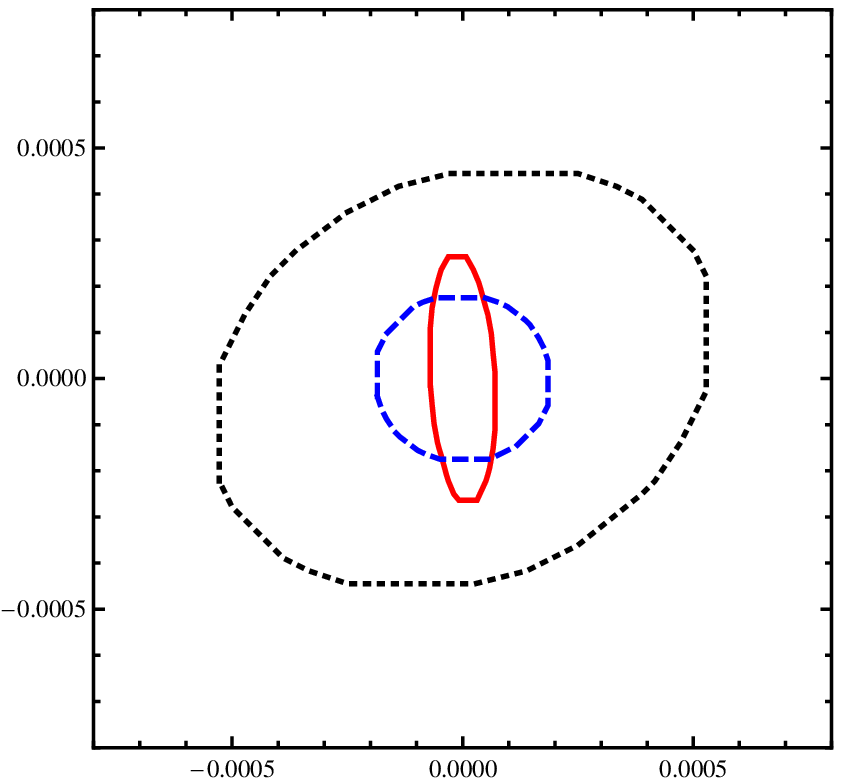,width=7.25cm}
\put(-335,-10){$\phi_2(\pi)$} \put(-105,-10){$\phi_2(\pi)$}
\put(-450,180){$\phi_3(\pi)$} \put(-215,180){$\phi_3(\pi)$}
\put(-340,175){tan$\beta=3$} \put(-115,175){tan$\beta=60$}\\
\epsfig{file=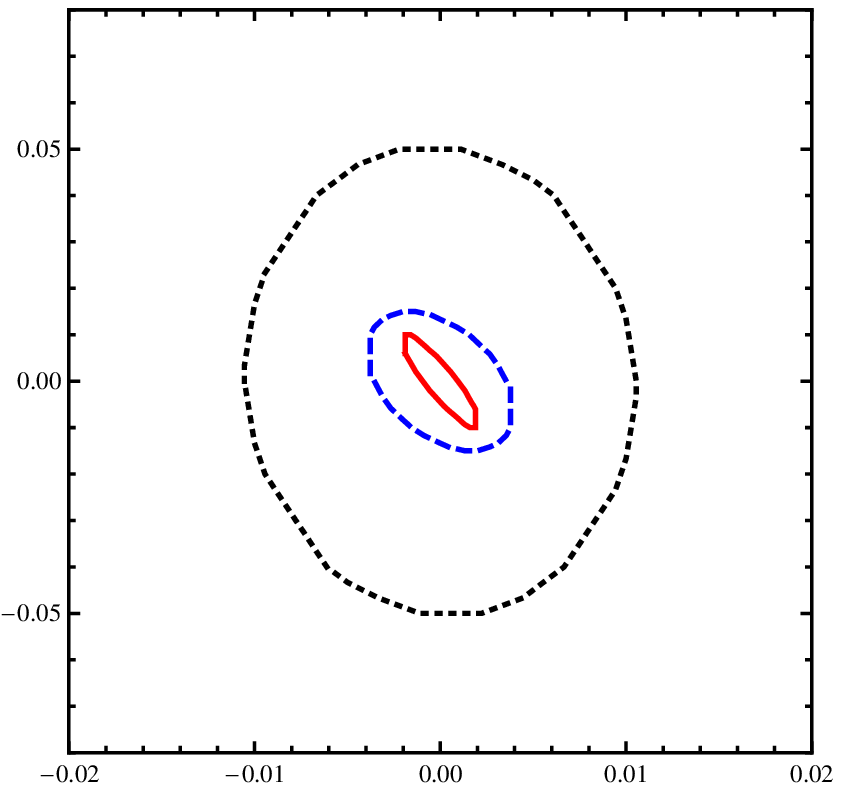,width=7.25cm}~~~~
\epsfig{file=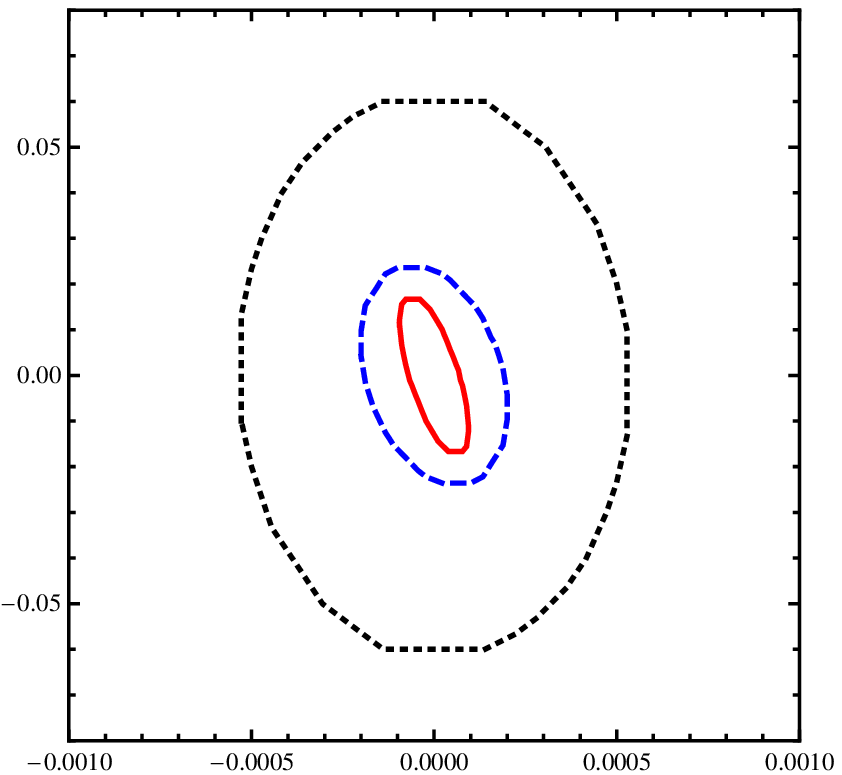,width=7.25cm}
\put(-335,-10){$\phi_2(\pi)$} \put(-105,-10){$\phi_2(\pi)$}
\put(-450,180){$\phi_{u,d}(\pi)$} \put(-225,180){$\phi_{u,d}(\pi)$}
\put(-340,175){tan$\beta=3$} \put(-115,175){tan$\beta=60$}\\
\epsfig{file=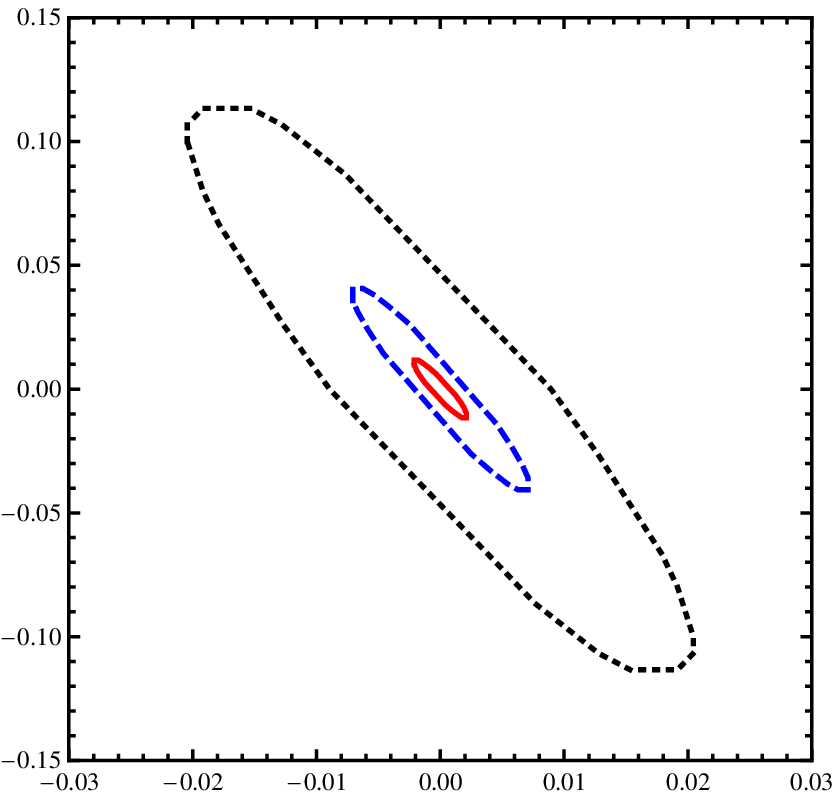,width=7.25cm}~~~~~
\epsfig{file=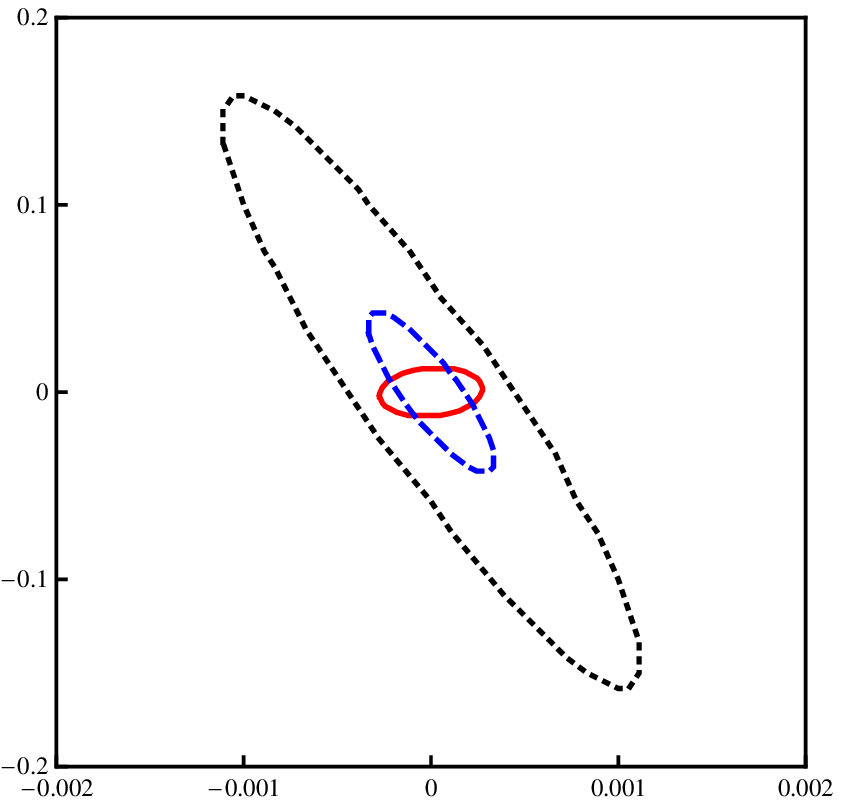,width=7.25cm}
\put(-335,-10){$\phi_3(\pi)$} \put(-105,-10){$\phi_3(\pi)$}
\put(-455,180){$\phi_{u,d}(\pi)$} \put(-228,180){$\phi_{u,d}(\pi)$}
\put(-340,175){tan$\beta=3$} \put(-115,175){tan$\beta=60$}
\end{center}

\caption{The correlated constraints on $\phi_2$ and $\phi_3$ (upper panels), $\phi_{u,d}$ (middle panels) and on $\phi_3$ and $\phi_{u,d}$ (lower panels), for
${\rm tan}\beta=3$, $60$ and $(m_{L,R})_1=200$ GeV (red solid line),
$500$ GeV (blue dashed line), and $1000$ GeV (black dotted line).
Points inside the curve satisfy the 95\% c.l. of the neutron,
Thallium, and Mercury EDM bounds.} \label{fig:phi23}
\end{figure}

We summarize how each phase is constrained by current EDM bounds in
Table \ref{tab:phase2edm}. The table shows that more than one phase
is strongly constrained in case I (light first generation
sfermions). These phases include $\phi_{2,3}$ in the
higgsino-gaugino sector, and $\phi_{u,d}$ in the squark sector. In
the following, we study the EDM constraints on the parameter space
defined by  pairs of such CP violating phases. For
simplicity, we tie $\phi_{u,d}$ together, but keep $\phi_2$ and
$\phi_3$ as independent. For each of the three combinations of
phases ($\phi_2$, $\phi_3$), ($\phi_2$, $\phi_{u,d}$), and
($\phi_3$, $\phi_{u,d}$), we perform $\chi^2$ analysis and determine the
region satisfying the combined neutron, Thallium, and
Mercury EDM bounds at 95\% c.l., for ${\rm tan}\beta=3$ and $60$, and
$(m_{L,R})_1=200,\ 500$ and  $1000$ GeV.

As shown in Fig. \ref{fig:phi23}, the allowed region for phases grows with the
increase in mass scales in most cases. The only exception is the
$\phi_3$ with large ${\rm tan}\beta$ (lower right panel of Fig.
\ref{fig:phi23}), which is due to the
non-monotonic behavior shown in Fig. \ref{fig:caseI}, lower panels.

The correlation between constraints on $\phi_3$ and $\phi_{u,d}$
(lower panels of Fig. \ref{fig:phi23}) is the strongest, since both of them are
dominantly constrained by the same bound, namely that from the Mercury EDM. On the
other hand, the correlation between ($\phi_2$, $\phi_3$) (Fig.
\ref{fig:phi23} upper panels) and ($\phi_2$, $\phi_{u,d}$) (Fig.
\ref{fig:phi23} middle panels) are rather weak, because $\phi_2$ is constrained
by different bounds -- neutron and Thallium EDMs.

\begin{table}[!b]
\caption{Summary of the combined bounds at 95\% c.l. on three phases
($\phi_2$, $\phi_3$, $\phi_{u,d}$) for ${\rm tan}\beta=3$, $60$ and
$(m_{L,R})_1=200$, $500$, and $1000$ GeV, using current experimental
limits of neutron, Thallium, and Mercury EDMs as in Eq. (1).}
\begin{center}
\label{tab:current}
\begin{tabular}{|c|c|c|c|c|c|c|}
\hline ${\rm tan}\beta$ &
\multicolumn{3}{|c|}{3} & \multicolumn{3}{|c|}{60} \\
\hline $(m_{L,R})_1$ &  200 GeV & 500 GeV & 1000 GeV & 200 GeV &
500 GeV & 1000 GeV \\
\hline $|\phi_2|$ & $<2.1\times 10^{-3}$ & $<5.0\times 10^{-3}$ &
$<1.5\times 10^{-2}$
& $<9.3\times 10^{-5}$ & $<2.5\times 10^{-4}$ & $<6.9\times 10^{-4}$ \\
\hline $|\phi_3|$ & $<2.8\times 10^{-3}$ & $<9.7\times 10^{-3}$ &
$<2.8\times 10^{-2}$
& $<3.1\times 10^{-4}$ & $<4.2\times 10^{-4}$ & $<1.5\times 10^{-3}$ \\
\hline $|\phi_{u,d}|$ & $<1.8\times 10^{-2}$ & $<6.0\times 10^{-2}$
& $<0.17$
& $<1.7\times 10^{-2}$ & $<5.6\times 10^{-2}$ & $<0.21$ \\

\hline
\end{tabular}
\end{center}
\end{table}

\begin{table}[!b]
\caption{Summary of the combined bounds at 95\% c.l. on three phases
($\phi_2$, $\phi_3$, $\phi_{u,d}$) for ${\rm tan}\beta=3$, $60$ and
$(m_{L,R})_1=200$, $500$, and $1000$ GeV, using current experimental
limits of neutron, Thallium EDMs as in Eq. (1), and previous limits
of Mercury EDM $d^{{\rm previous}}_{{\rm Hg}} = 2\times 10^{-28} e
{\rm cm}$ (95\% c. l.)\cite{Romalis:2000mg}.}
\begin{center}
\label{tab:oldmercury}
\begin{tabular}{|c|c|c|c|c|c|c|}
\hline ${\rm tan}\beta$ &
\multicolumn{3}{|c|}{3} & \multicolumn{3}{|c|}{60} \\
\hline $(m_{L,R})_1$ &  200 GeV & 500 GeV & 1000 GeV & 200 GeV &
500 GeV & 1000 GeV \\
\hline $|\phi_2|$ & $<2.1\times 10^{-3}$ & $<5.0\times 10^{-3}$ &
$<1.5\times 10^{-2}$
& $<9.2\times 10^{-5}$ & $<2.5\times 10^{-4}$ & $<6.9\times 10^{-4}$ \\
\hline $|\phi_3|$ & $<8.7\times 10^{-3}$ & $<2.3\times 10^{-2}$ &
$<6.1\times 10^{-2}$
& $<1.9\times 10^{-3}$ & $<1.5\times 10^{-3}$ & $<3.6\times 10^{-3}$ \\
\hline $|\phi_{u,d}|$ & $<2.3\times 10^{-2}$ & $<6.6\times 10^{-2}$
& $<0.18$
& $<2.8\times 10^{-2}$ & $<5.6\times 10^{-2}$ & $<0.23$ \\

\hline
\end{tabular}
\end{center}
\end{table}

\begin{table}[!b]
\caption{Summary of the combined bounds at 95\% c.l. on three phases
($\phi_2$, $\phi_3$, $\phi_{u,d}$) for ${\rm tan}\beta=3$, $60$ and
$(m_{L,R})_1=200$, $500$, and $1000$ GeV, using current experimental
limits of Thallium and Mercury EDMs as in Eq. (1), and future
possible improved neutron EDM limit $d^{{\rm future}}_{{\rm n}} =
2.9\times 10^{-28} e {\rm cm}$ (90\% c. l.).}
\begin{center}
\label{tab:futureneutron}
\begin{tabular}{|c|c|c|c|c|c|c|}
\hline ${\rm tan}\beta$ &
\multicolumn{3}{|c|}{3} & \multicolumn{3}{|c|}{60} \\
\hline $(m_{L,R})_1$ &  200 GeV & 500 GeV & 1000 GeV & 200 GeV &
500 GeV & 1000 GeV \\
\hline $|\phi_2|$ & $<2.1\times 10^{-3}$ & $<4.4\times 10^{-3}$ &
$<1.3\times 10^{-2}$
& $<3.6\times 10^{-5}$ & $<2.2\times 10^{-4}$ & $<5.6\times 10^{-4}$ \\
\hline $|\phi_3|$ & $<7.4\times 10^{-4}$ & $<6.1\times 10^{-3}$ &
$<1.9\times 10^{-2}$
& $<2.8\times 10^{-4}$ & $<2.8\times 10^{-4}$ & $<7.4\times 10^{-4}$ \\
\hline $|\phi_{u,d}|$ & $<1.2\times 10^{-2}$ & $<3.0\times 10^{-2}$
& $<7.8\times 10^{-2}$
& $<6.0\times 10^{-3}$ & $<2.2\times 10^{-2}$ & $<7.6 \times 10^{-2}$ \\

\hline
\end{tabular}
\end{center}
\end{table}

\begin{table}[!b]
\caption{Summary of the combined bounds at 95\% c.l. on three phases
($\phi_2$, $\phi_3$, $\phi_{u,d}$) for ${\rm tan}\beta=3$, $60$ and
$(m_{L,R})_1=200$, $500$, and $1000$ GeV, using current experimental
limits of neutron and Mercury EDMs as in Eq. (1), and future
possible improved Thallium EDM limit $d^{{\rm future}}_{{\rm Tl}} =
9.0\times 10^{-27} e {\rm cm}$ (90\% c. l.).}
\begin{center}
\label{tab:futureThallium}
\begin{tabular}{|c|c|c|c|c|c|c|}
\hline ${\rm tan}\beta$ &
\multicolumn{3}{|c|}{3} & \multicolumn{3}{|c|}{60} \\
\hline $(m_{L,R})_1$ &  200 GeV & 500 GeV & 1000 GeV & 200 GeV &
500 GeV & 1000 GeV \\
\hline $|\phi_2|$ & $<2.1\times 10^{-5}$ & $<5.2\times 10^{-5}$ &
$<1.7\times 10^{-4}$
& $<9.3\times 10^{-7}$ & $<2.5\times 10^{-6}$ & $<7.6\times 10^{-6}$ \\
\hline $|\phi_3|$ & $<2.8\times 10^{-3}$ & $<7.8\times 10^{-3}$ &
$<2.2\times 10^{-2}$
& $<3.1\times 10^{-4}$ & $<3.8\times 10^{-4}$ & $<1.2\times 10^{-3}$ \\
\hline $|\phi_{u,d}|$ & $<1.5\times 10^{-2}$ & $<4.6\times 10^{-2}$
& $<0.13$
& $<1.5\times 10^{-2}$ & $<5.0\times 10^{-2}$ & $<0.19$ \\

\hline
\end{tabular}
\end{center}
\end{table}

As a final illustration of the impact of the EDM constraints, we show in Table \ref{tab:current} the combined bounds (at
95\% c.l.) on all the three phases ($\phi_2$, $\phi_3$,
$\phi_{u,d}$) implied by the current EDM experimental limits. For  purposes
of comparison, we also show the impact of (a) the reduction in the Mercury EDM bound compared to the previous result ($d^{{\rm previous}}_{{\rm Hg}} = 2\times 10^{-28} e {\rm cm}$(95\%
c. l.)) ( see Table \ref{tab:oldmercury});
(b) a neutron EDM bound that is 100 times more stringent than the present limit, keeping all other EDM limits as at present (\ref{tab:futureneutron}); and (c) a similar improvement in the Thallium limit, keeping the neutron and Mercury limits as at present (\ref{tab:futureThallium}).

The bounds on phases ($\phi_2$, $\phi_3$, $\phi_{u,d}$) from the
combined 3-phase $\chi^2$ analysis, as shown in Table
\ref{tab:current}, are in fact rather close to the bounds from the
combined 2-phase $\chi^2$ analysis, for $\phi_2$ shown in the upper and middle panels of Fig.
\ref{fig:phi23}, and for $\phi_3$ and
$\phi_{u,d}$ shown in the lower panel of Fig. \ref{fig:phi23}, simply because a
strong correlation only exists between $\phi_3$ and $\phi_{u,d}$,
and because the inclusion of $\phi_2$ does not substantially alter this correlation.

Comparing Table \ref{tab:current} and \ref{tab:oldmercury}, we
observe that the most significant impact of the recent update on Mercury
EDM bound is on $\phi_3$, while there is no impact on $\phi_2$ and
only small impact on $\phi_{u,d}$. This can be understood by looking
at Fig. \ref{fig:caseI} (middle and lower panels), \ref{fig:phiu}, and
\ref{fig:phid}. The most stringent bounds on $\phi_2$ are from the
Thallium and neutron EDMs. The most stringent bound on $\phi_3$ is
from the current Mercury EDM, and the second most stringent bound
from neutron EDM is many times looser. The most stringent bound on
$\phi_d$ is from the Mercury EDM bound, but the neutron EDM bound is
rather close.

In principle, one might expect that the impact of future neutron EDM experiments would be apparent when
comparing Table \ref{tab:current} and \ref{tab:futureneutron}.
Assuming the neutron EDM limit becomes 100 times tighter than
current one, the limits on all phases $\phi_2$, $\phi_3$, and
$\phi_{u,d}$, however, would only change to be, at most, a few times
smaller. This is because  as the neutron EDM bound becomes the most
stringent one for $\phi_2$, $\phi_3$, and $\phi_{u,d}$, 
 strong correlations among all of these three phases emerge, making the
limits on all of them much larger than when the we consider limits on individual phases. This correlation arises from the presence of cancellations between various contributions associated with the different phases, a situation that has been noted previously in the literature (see, {\em e.g.}, Ref.~\cite{Ibrahim:1998je}).

\begin{figure}
\begin{center}
\epsfig{file=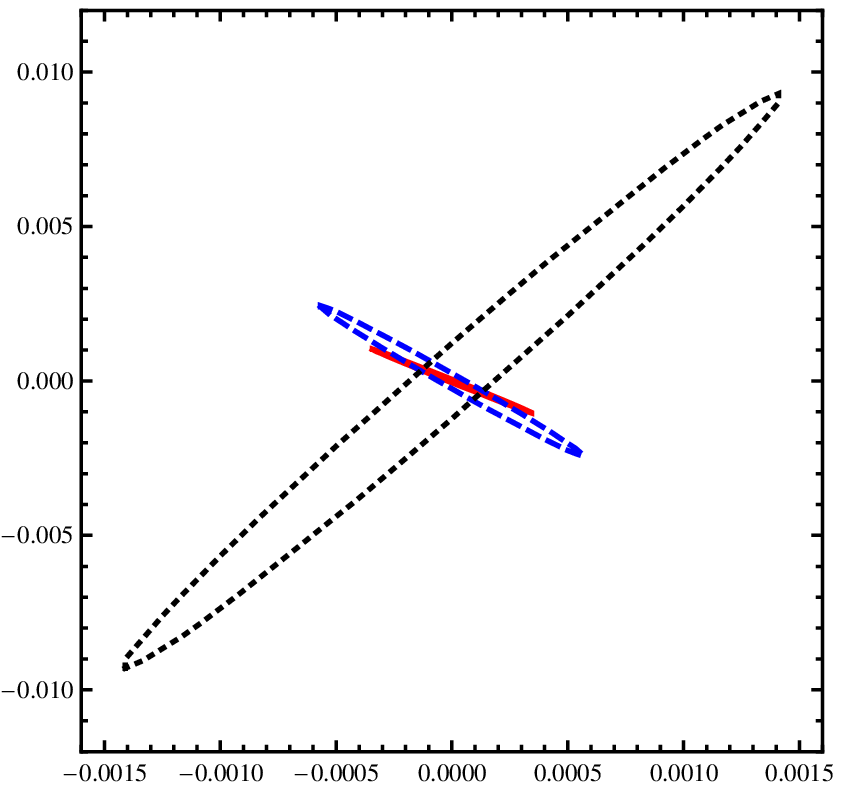,width=7.25cm}~~~~
\epsfig{file=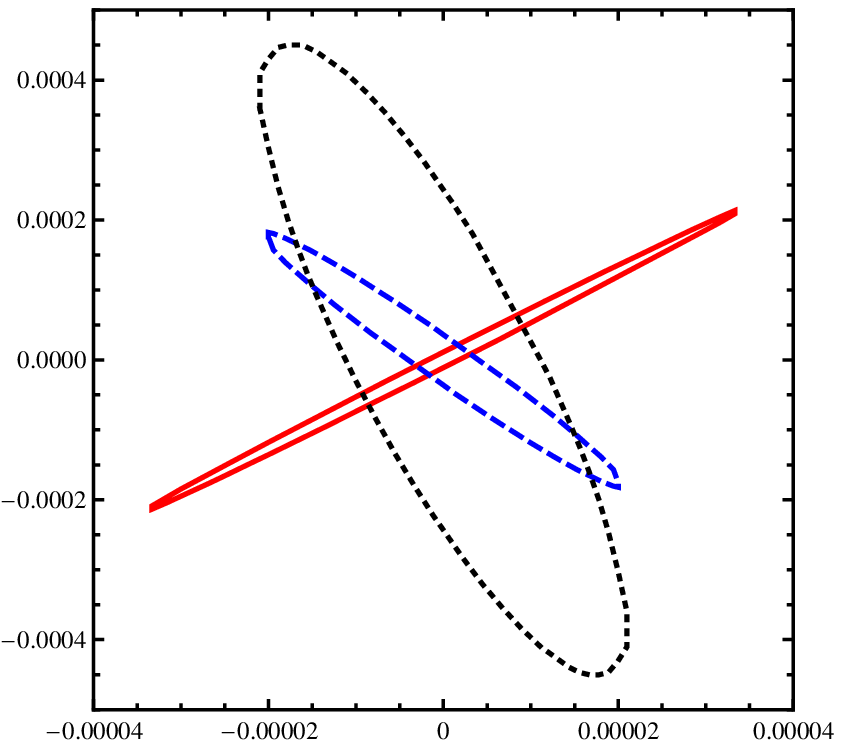,width=7.7cm}
\put(-345,-10){$\phi_2(\pi)$} \put(-115,-10){$\phi_2(\pi)$}
\put(-460,180){$\phi_3(\pi)$} \put(-225,180){$\phi_3(\pi)$}
\put(-340,175){tan$\beta=3$} \put(-115,175){tan$\beta=60$}\\
\epsfig{file=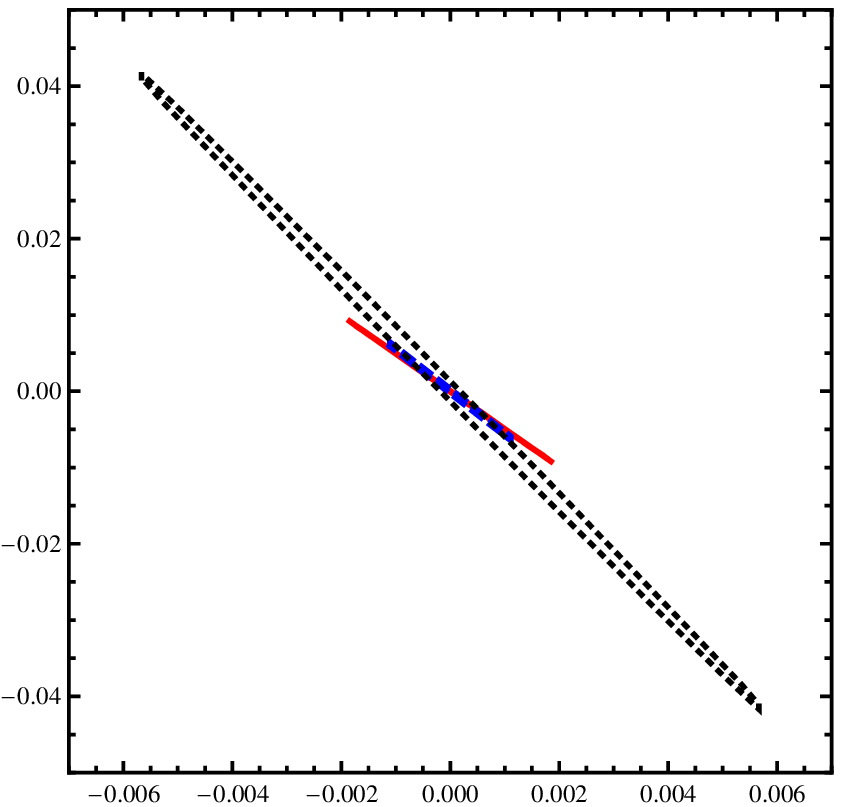,width=7.25cm}~~~~~~
\epsfig{file=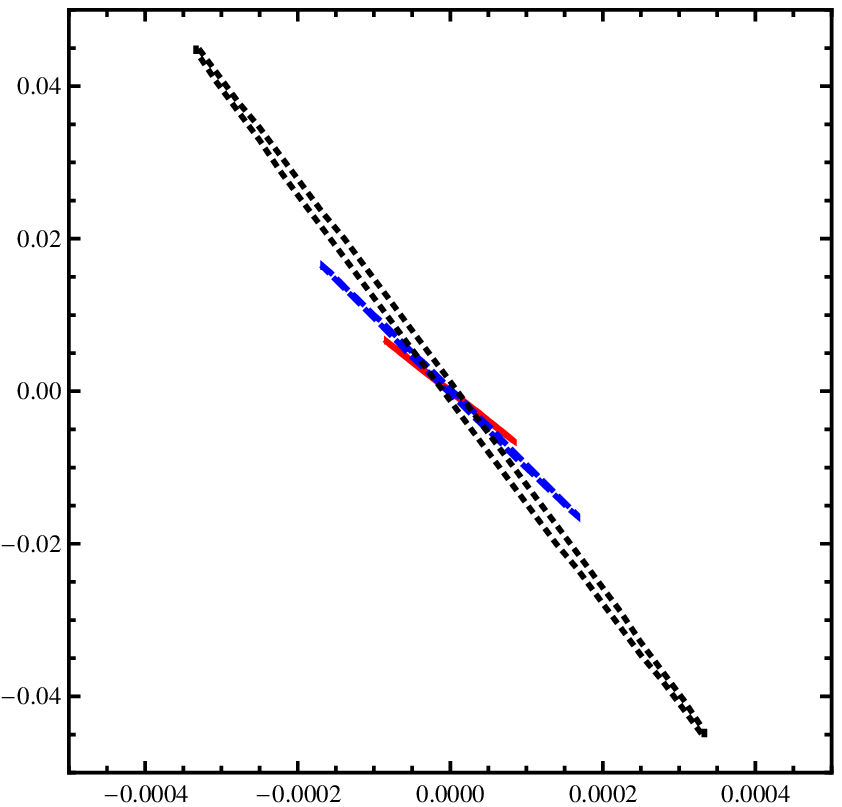,width=7.25cm}
\put(-335,-10){$\phi_2(\pi)$} \put(-105,-10){$\phi_2(\pi)$}
\put(-455,185){$\phi_{u,d}(\pi)$} \put(-225,185){$\phi_{u,d}(\pi)$}
\put(-340,175){tan$\beta=3$} \put(-115,175){tan$\beta=60$}\\
\epsfig{file=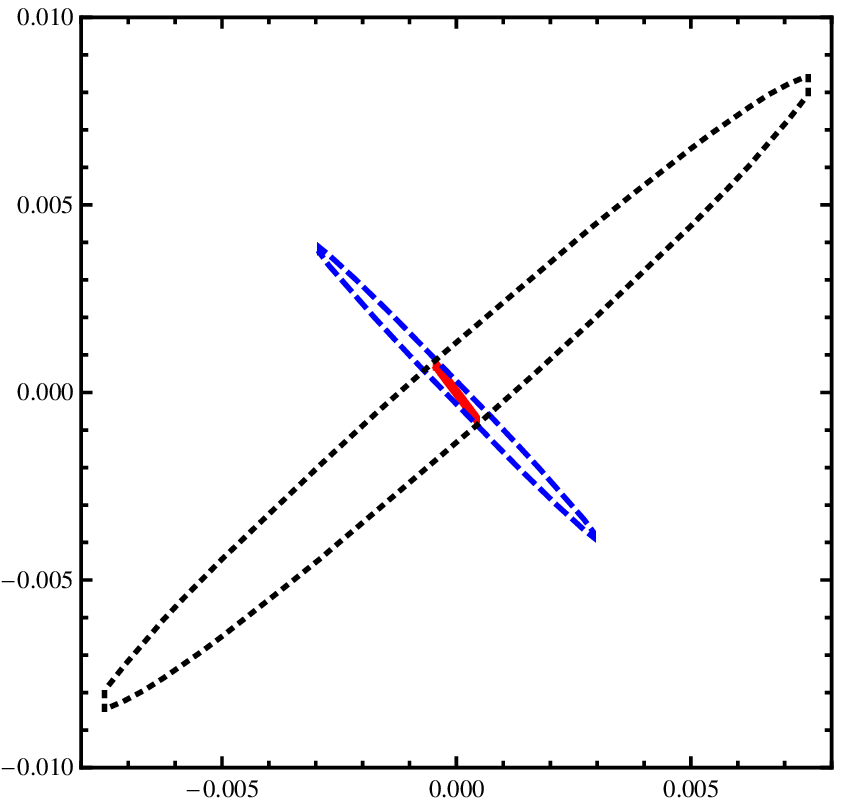,width=6.9cm}~~~~~~
\epsfig{file=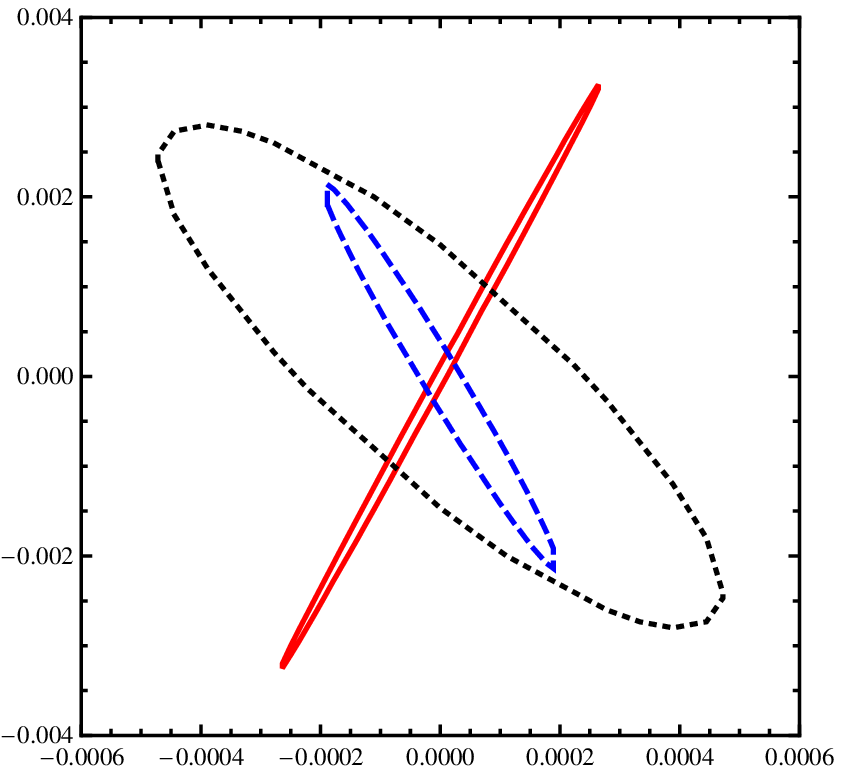,width=7.25cm}
\put(-335,-10){$\phi_3(\pi)$} \put(-105,-10){$\phi_3(\pi)$}
\put(-455,175){$\phi_{u,d}(\pi)$} \put(-225,175){$\phi_{u,d}(\pi)$}
\put(-340,175){tan$\beta=3$} \put(-115,175){tan$\beta=60$}
\end{center}

\caption{The correlated constraints on $\phi_2$ and $\phi_3$ (upper panels), $\phi_{u,d}$ (middle panels) and on $\phi_3$ and $\phi_{u,d}$ (lower panels), for
${\rm tan}\beta=3$, $60$ and $(m_{L,R})_1=200$ GeV (red solid line),
$500$ GeV (blue dashed line), and $1000$ GeV (black dotted line).
Points inside the curve satisfy the 95\% c.l. of the
future neutron EDM bound,
and current Thallium and Mercury EDM bounds.} \label{fig:newdnphi23}
\end{figure}

The correlations arising from this \lq\lq cancellation mechanism" is most easily observed by considering the combined constraints on pairs of phases.
To illustrate, 
we show in Fig. \ref{fig:newdnphi23},
 the correlated constraints on ($\phi_2$,
$\phi_3$), ($\phi_2$,
$\phi_{u,d}$), and ($\phi_3$, $\phi_{u,d}$), respectively, using a prospective future neutron EDM bound.
Comparing with Fig. \ref{fig:phi23}, we see that with a future neutron EDM bound that is 100 times tighter, the
neutron EDM would become the dominant constraint. In such a case,
the cancellation exists among all the phases $\phi_2$,
$\phi_3$, and $\phi_{u,d}$, leading to a much narrower region for allowed phase values. We also note that the orientations of the 95 \% C.L. ellipses involving $\phi_3$ can differ from what appears in Fig. \ref{fig:phi23} since for the neutron, the constraints arise from the effects of {\em both} the quark EDMs and chromo-EDMs, in contrast to the situation for Mercury where the chromo-EDMs dominate. The relative importance of the $d_q$ and $\tilde{d}_q$ contributions to $d_n$ can change with $m_{L,R}$, leading to changes in the orientation of the ellipses with the value of these mass parameters.

To asses the impact of the future Thallium EDM experiments, we
compare Table \ref{tab:current} with \ref{tab:futureThallium}.
Assuming the future Thallium EDM bound is 100 times tighter than
current one, the limit on $\phi_2$ would correspondingly shrink by
approximately 100 times, while the limit on $\phi_3$ and
$\phi_{u,d}$ would not change much, indicating a 100 times tighter
Thallium EDM bound would not be more stringent than current neutron
and Mercury EDM bounds on constraining $\phi_3$ and $\phi_{u,d}$.

\subsection{Implication of loosely constrained phases on CP-violating and CP-conserving phenomenologies.}

As the EDM sensitivities improve, the impact of the most strongly constrained phases on other observables will be reduced. On the other hand, the effects of the more loosely constrained phases may still be apparent. 
In the higgsino-gaugino sector, for example, the only loosely constrained phase is
$\phi_1$ (for relatively heavy first generation sfermions and small to moderate $\tan\beta$). This has important consequences in a rather wide range of
phenomenological contexts, the most significant one being electroweak baryogenesis. A large $\phi_1$ can generate a sizeable contribution to the 
baryon asymmetry via bino-driven electroweak baryogenesis scenario,
while the other scenario (wino-driven), which depends on $\phi_2$,
is highly constrained by current EDM bounds \cite{Li:2008ez}.

Moreover, this phase is in general present in processes involving
neutralinos both on-shell and off-shell. Its effect shifts the neutralino
mass spectrum and modify the couplings. Among CP-conserving
quantities, it can lead to  order one changes in the production rate,
decay width, and branching ratios of neutralinos at colliders
\cite{Choi:1999cc,Barger:2001nu}. It also modifies the relic density, as
well as direct and indirect detection rate of neutralino dark matter
\cite{phi1DM}. For example, as shown in Ref. \cite{phi1DM}, the
typical variations in the neutralino relic abundance from the $\phi_1$-dependence of couplings is about ${\cal O}(10 -100\%$).

A potentially more direct probe of $\phi_1$ may be through its impact on CP-violating
observables at both the LHC and a future linear collider. At the
LHC, it contributes to the triple product
$(\vec{p}_1\times\vec{p}_2)\cdot\vec{p}_3$ associated with cascade
decays of stops \cite{Bartl:2004jr,Langacker:2007ur,Ellis:2008hq}
\begin{equation}
gg \rightarrow \tilde{t}_i \tilde{t}_i, ~~~ \tilde{t}_i \rightarrow
t \chi^0_j, ~~~  \chi^0_j \rightarrow \chi^0_1 l^+ l^-,
\end{equation}
where the $\vec{p}_i$ are the momenta of final state charged particles in the decay chain of the
stop. This observable is manifestly T-odd. If strong phases are negligible, then it can provide a probe of CPV.
 It is shown in Ref.
\cite{Langacker:2007ur} that, assuming an order one phase, the signal
can be detected with $10^2 - 10^3$ identified events. At a future
linear collider, it may contribute to the triple product
$(\vec{p}_{l^+}\times\vec{p}_{l^-})\cdot\vec{p}_{e^+}$ associated
with neutralino production and subsequent leptonic decays
\cite{Choi:1999cc,Barger:2001nu,Kizukuri:1990iy,Bartl:2003tr,Bartl:2004jj}
\begin{equation}
e^+ e^- \rightarrow \chi^0_1 \chi^0_2, ~~~ \chi^0_2 \rightarrow
\chi^0_1 l^+ l^-,
\end{equation}
which is a genuine CP-odd observable. As shown in Ref.
\cite{Bartl:2004jj}, the CP asymmetry in this process can reach
$10\%$ for some values of the mass and phase parameters. It also
contributes to another triple-product
$(\vec{p}_{\tau}\times\vec{p}_{e^+})\cdot\vec{s}_{\tau}$, a T-odd
observable that is constructed using the transverse polarization
$\vec{s}_{\tau}$ of the $\tau^{\pm}$ in the neutralino two-body
decay
\begin{equation}
e^+ e^- \rightarrow \chi^0_1 \chi^0_2, ~~~ \chi^0_2 \rightarrow
\tilde{\tau}^{\pm} \tau^{\mp}. \label{eq:tautau}
\end{equation}
This correlation has been studied in Ref. \cite{CPVILCtau}, where it is shown
that the corresponding asymmetry can reach values up to $60\%$.

Finally, an off-shell neutralino in loop can in principle generate
CP-odd observables in B-meson decays. However, the neutralino
contributions is in most cases subdominant compared to other
contributions involving gluions and charginos \cite{Bphysics}.

Among all the loosely constrained phases $\phi_{e,\mu,\tau}$ in the
slepton sector and $\phi_{c,s,t,b}$ in the squark sector, the
implication of the third-generation phases $\phi_{\tau}$ and
$\phi_{t,b}$ are most interesting. The phase $\phi_{\tau}$
contributes to the aforementioned T-odd observable
$(\vec{p}_{\tau}\times\vec{p}_{e^+})\cdot\vec{s}_{\tau}$ in the
process outlined in Eq. (\ref{eq:tautau}) \cite{CPVILCtau}. The phase
$\phi_{t,b}$ may generate sizable effect in B-meson physics, even
though they are flavor-conserving by themselves. The
flavor-violation can be either from within the SM (CKM quark
mixing), or from beyond the SM (off-diagonal elements in squark mass
matrices). For example, a large $\phi_t$ can generate sizable
deviation from SM for $S_{\phi(\eta'){K_S}}$, and generate large CP
asymmetries $A_{CP}(b \rightarrow s \gamma)$ in $b \rightarrow s
\gamma$, through contributions involving charginos and
charged-Higgs, where the flavor-violation comes from CKM mixing
matrix \cite{flavorblind}. On the other hand, a deviation from SM
for $S_{\phi{K_S}}$ can also be generated by chirality-flipping $LR$
and $RL$ gluino contributions \cite{Kane:2003zi}, which can be
induced by a flavor-violating chirality-conserving mass insertion
$(\delta^d_{LL,RR})_{23}$ and a chirality-flipping CP-violating mass
insertion associated with the $\phi_b$: $(\delta^d_{LR,RL})^{{\rm
induced}}_{23}=(\delta^d_{LL,RR})_{23} \times m_b(A_b-\mu {\rm
tan}\beta)/\tilde{m}^2$.

The phases $\phi_{t,b}$ also generate mixing between CP-even and
CP-odd Higgses in MSSM, changing their mass spectrum and couplings
\cite{Pilaftsis:1999qt}, which has important consequences for 
Higgs searches at the Tevatron and LHC \cite{Carena:2002bb}. As another
example of their impact on CP-conserving observables, the large
$\phi_{t,b}$ also changes the neutralino annihilation and scattering
cross section and thus are important for relic density, as well as for
direct and indirect detection rates for neutralino dark matter
\cite{phitbDM}.

\section{Conclusions}
\label{sec:conclusions} In the present study, we analyzed the
constraints from electric dipole moments on the size of CP-violating
phases in the MSSM, utilizing the dominant one- and two-loop
contributions. We introduced the {\tt 2LEDM} numerical code, interfaced to CPSuperH2.0, that encompasses all these contributions. We pointed out that not all CP violating phases in
the MSSM are constrained to be small by null results from EDM
searches. Our results are summarized for the ease of the reader in
Table \ref{tab:phase2edm}. We differentiate there between a case
where the first generation sfermions are light, and one-loop
contributions to EDMs are significant, and one where they are heavy
and one-loop contributions to EDMs are consequently suppressed.

We find that in the gaugino sector, the ``wino'' phase $\phi_2$ is
fairly strongly constrained in both cases of light and heavy first
generation sfermion masses $(m_{L,R})_1$, the ``gluino'' phase
$\phi_3$ is only strongly constrained in the case with light
$(m_{L,R})_1$, and the ``bino'' phase $\phi_1$ can be arbitrarily
large in the case with heavy $(m_{L,R})_1$, and can be sizable with
light $(m_{L,R})_1$ and small ${\rm tan}\beta$. We note that in earlier studies that employed a universality 
assumption ($\phi_1=\phi_2=\phi_3$) this difference in EDM sensitivities to the different CPV phases in the gauge-Higgs sector of the MSSM was not apparent. 

Turning to sfermion CP violating phases, only the stop CP violating
phase $\phi_t$ (and only more weakly $\phi_b$) is marginally
constrained in the limits of heavy first generation squarks and
sleptons. For light first generation sfermions, rather stringent
constraints arise for $\phi_d$ and for $\phi_u$ from the neutron and
Mercury EDM bounds, and weak constraint for $\phi_e$ arises from the
Thallium EDM limit. Virtually no constraint exists from EDMs on the
CP violating phases for second and third generation leptons and for
second generation squarks.

In summary, we showed in this paper that even in the absence of
cancellations between contributions from different CP violating
phases, null results on searches for the permanent EDM of the
neutron and of atoms put constraints only on selected CP violating
phases, leaving ample room for a rich phenomenology related to CP
violation at colliders, B factories and dark matter searches.

\section*{Acknowledgements}
M.J.R-M. thanks I. Rothstein for pointing out the recent CDF results. Y. Li  thanks Xiangdong Ji for his hospitality at Shanghai Jiaotong University, where part of this work was completed
S.P. is partly supported by an Outstanding Junior Investigator Award
from the US Department of Energy (DoE), Office of Science, High
Energy Physics, and by DoE Contract DE-FG02-04ER41268, NSF Grant
PHY-0757911 and a Faculty Research Grant from the University of
California, Santa Cruz. Y. Li and M.J.R-M. were supported in part by DoE contract DE-FG02-08ER41531 and by the Wisconsin Alumni Research Foundation.

\end{document}